\documentclass[12pt,times]{article}
 \usepackage[pdftex]{color,graphicx}
\usepackage{enumerate}
\usepackage{amsmath}
\usepackage{amssymb}
\usepackage{alltt}
\usepackage{setspace}
\usepackage{subfigure}
\usepackage{sidecap}
\usepackage{sectsty}
\partfont{\large}
\sectionfont{\large}
\subsectionfont{\small}
\usepackage[left=2cm,top=2cm,right=3cm,head=1cm,foot=1cm]{geometry}
\usepackage{mathrsfs}
\numberwithin{equation}{section}
\let\oldsqrt\sqrt
\def\sqrt{\mathpalette\DHLhksqrt}
\def\DHLhksqrt#1#2{%
\setbox0=\hbox{$#1\oldsqrt{#2\,}$}\dimen0=\ht0
\advance\dimen0-0.2\ht0
\setbox2=\hbox{\vrule height\ht0 depth -\dimen0}%
{\box0\lower0.4pt\box2}}
\newcommand{\al}{\alpha}
\newcommand{\g}{\gamma}
\newcommand{\e}{\varepsilon}
\newcommand{\ta}{\theta}

\newcommand{\G}{\Gamma}

\newcommand{\ph}{\varphi}
\newcommand{\da}{\dagger}
\newcommand{\pa}{\partial}

\newcommand{\la}{\mathscr{L}}
\newcommand{\ld}{\lambda}

\newcommand{\x}{\times}
\newcommand{\ox}{\otimes}

\newcommand{\ml}{\left(\begin{matrix}}
\newcommand{\mr}{\end{matrix}\right)}

\newcommand{\bra}{\langle}
\newcommand{\ket}{\rangle}
\newcommand{\tr}{\text{tr}}
\newcommand{\op}{\mathcal O}
\newcommand{\del}{\delta}

\newcommand{\zb}{\mathbb Z}

\newcommand{\half}{\tfrac{1}{2}}
\newcommand{\third}{\tfrac{1}{3}}
\newcommand{\fourth}{\tfrac{1}{4}}
\newcommand{\sixth}{\tfrac{1}{6}}
\newcommand{\eighth}{\tfrac{1}{8}}

\newcommand{\s}{\sigma}

\newcommand{\sq}{\tfrac{1}{\sqrt2}}

\newcommand{\Ds}{\mathscr D}

\newcommand{\As}{\mathscr A}
\newcommand{\Bs}{\mathscr B}

\newcommand{\Ns}{\mathscr N}

\newcommand{\Ta}{\Theta}

\newcommand{\intsum}{\int \!\!\!\!\!\!\!\!\sum}

\begin{document}
\title{Fermion masses without symmetry breaking in two spacetime dimensions}
\date{}
\author{Yoni BenTov \\ \small Department of Physics, University of California, Santa Barbara, CA 93106, USA}
\maketitle
\abstract{I study the prospect of generating mass for symmetry-protected fermions without breaking the symmetry that forbids quadratic mass terms in the Lagrangian. I focus on 1+1 spacetime dimensions in the hope that this can provide guidance for interacting fermions in 3+1 dimensions. I first review the $SO(8)$ Gross-Neveu model and emphasize a subtlety in the triality transformation. Then I focus on the ``$m = 0$" manifold of the $SO(7)$ Kitaev-Fidkowski model. I argue that this theory exhibits a phenomenon similar to ``parity doubling" in hadronic physics, and this leads to the conclusion that the fermion propagator vanishes when $p^{\,\mu} = 0$. I also briefly explore a connection between this model and the two-channel, single-impurity Kondo effect. This paper may serve as an introduction to topological superconductors for high energy theorists, and perhaps as a taste of elementary particle physics for condensed matter theorists.}
\section{Introduction}\label{sec:intro}
In an effort to demonstrate the interdisciplinary value of the study of topological superconductors, let me begin with a problem in elementary particle physics in 3+1 spacetime dimensions. In the Standard Model (SM) of particle physics, all fundamental fermions are massless at energies above the electroweak scale, $v = 246$ GeV. This is because one imposes the gauge symmetry
\begin{equation}
G_{\text{SM}} = SU(3)_{\text{color}} \x SU(2)_{\text{weak}} \x U(1)_{\text{hypercharge}}\; \nonumber
\end{equation}
and assigns a single generation of fermions to the representation:
\begin{equation}
(3,2,+\sixth)\oplus (\bar3,1,-\tfrac{2}{3})\oplus (\bar 3,1,+\third)\oplus(1,2,-\half)\oplus(1,1,+1)\;. \nonumber
\end{equation}
This representation is chiral and hence does not admit a $G_{\text{SM}}$-invariant mass term for any fundamental fermion field. The simplest way to give the fermions a mass at energies below the scale $v$ is to posit the existence of a spin-0 Higgs field transforming as $(1,2,-\half)$ and then to write a gauge-invariant Yukawa interaction. When the Higgs condenses, the electroweak part of the gauge group, $G_{\text{EW}} = SU(2)_{\text{weak}}\x U(1)_{\text{hypercharge}}$, is broken to $U(1)_{\text{EM}}$, and the fermions obtain mass.
\\\\
It is by now widely accepted that the quarks and leptons of the SM obtain masses in this way. The recent experimental discovery of the Higgs boson \cite{atlas, cms} strongly reinforces the expectation that the fermions should be massless in the $G_{\text{EW}}$-symmetric phase and massive in the $G_{\text{EW}}$-broken phase.
%
\\\\
As a matter of theoretical interest, it is worth emphasizing that the above picture is based on weak coupling perturbation theory. One might instead consider non-perturbatively large interactions and ask the following question: Is it possible for the fundamental fermions of the SM to obtain mass in the $G_{\text{EW}}$-symmetric phase?
\\\\
This is exactly the type of question that condensed matter theorists ask when they speak of ``reducing the classification of topological superconductors" \cite{KF, KF 2, ludwig gravitational response, modular, qi Z8, fidkowski Z16, fidkowski Z16 2, senthil 3d SPT}. It turns out that there are strong physical indications that, if one includes a gauge-singlet antineutrino per generation, then all physical excitations in the SM can be fully gapped without breaking any part of the SM gauge group \cite{wen so(10), cenke SM, cenke SM 2}. (The relationship between the reduced classification of topological superconductors and the anomaly matching condition was discussed in \cite{wang wen}.) It might be thought that the phenomenon studied in condensed matter physics is simply an artifact of the lattice and should not have a continuum description. However, there are recent numerical results which support the conjecture that the transitions in question are second order and should be described by an interacting quantum field theory \cite{mass without condensates 1, mass without condensates 2}. 
\\\\
This paper focuses on this type of problem in 1+1 spacetime dimensions within the framework of ``symmetry protected topological" (SPT) phases \cite{qi zhang TI, senthil SPT}. In this context, the symmetry group $G$ is a \textit{global} symmetry of the model, but it is often a useful theoretical device to gauge that symmetry by the usual minimal coupling procedure \cite{senthil 3d SPT}. 
\\\\
An SPT phase in $d$ spatial dimensions with global symmetry $G$ is a zero-temperature state of quantum matter whose three defining phenomenological properties are:
\begin{enumerate}[1)]
\item In a system without spatial boundaries, the ground state is unique and all excitations above the ground state are gapped. (The bulk is said to be ``trivial".)
\item In a system with spatial boundaries, the ground state is degenerate or there exist gapless excitations. (The boundary is said to be ``nontrivial".)
\item The boundary theory cannot be defined self-consistently as an independent quantum theory in $(d-1)$ spatial dimensions.
\end{enumerate}
If the global symmetry $G$ is broken (either spontaneously or by an explicit $G$-breaking term in the Lagrangian), then the formerly gapless boundary excitations become gapped, and the theory flows to a trivial gapped state at low energy. 
\\\\
The simplest field theoretic example is the continuum limit of the $1d$ Kitaev chain \cite{kitaev chain}. (For a review, see Appendix~\ref{sec:kitaev chain}.) Consider a (1+1)-dimensional relativistic theory of a massless Majorana fermion,\footnote{The spinor $\Ns$ satisfies the Majorana condition $\Ns^C \equiv C^{-1}\Ns^* = \Ns$ with $C = C^{-1} = C^T = \s_3$ when $\eta^* = \eta$ and $\bar\eta^* = \bar\eta$. My choice of gamma matrices is $\g^\mu = (\s_1,-i\s_2)$ and $\g^5 = \g^0\g^1 = \s_3$. It is more conventional to choose $C = I$, but I prefer $C = \s_3$ because that is compatible with an extension to 2+1 dimensions (with $\g^2 = i\g^5$). Alternatively, one could choose the ``Majorana basis" with $\g^\mu = (\s_2,i\s_1)$, in which case $C = I$ would also work for 2+1 dimensions.}
\begin{equation}
\Ns = \ml \bar\eta\\ i\eta \mr\;,
\end{equation}
coupled to a time-independent, spatially-dependent, semiclassical background scalar field $\phi(x)$ (here $x$ stands for the spatial coordinate only). The Lagrangian is:
\begin{equation}\label{eq:majorana lagrangian}
\la = \half \bar\Ns \left[i\!\!\not\!\pa-g\,\phi(x)\right]\Ns\;.
\end{equation}
Consider the infinitesimally thin kink profile:
\begin{equation}
g\,\phi(x) = \left\{  \begin{matrix} +m\;,\;\; x > 0\\ 0\;,\;\; x = 0\\ -m\;,\;\; x < 0 \end{matrix}\right.\;.
\end{equation}
For $x > 0$ there is a free Majorana fermion with a physical mass $m$, and for $x < 0$ there is also a free Majorana fermion with physical mass $m$. (By the value $m$ being the ``physical mass" I mean that the fermion transforms as the Poincar\'e representation $p^2 = -m^2$.) 
\\\\
However, at $x = 0$ there is a time-independent real fermion stuck to the core of the kink. To see this \cite{zero modes}, write $\Ns = \Ns_+ + \Ns_-$, where $\g^5\Ns_\pm = \pm \Ns_\pm$. The equations of motion $\del\la/\del\bar\Ns_+ = 0$ and $\del\la/\del\bar\Ns_- = 0$ admit a solution of the form 
\begin{equation}
\Ns_+ = \ml c\\ 0 \mr e^{-m|x|}\;,\;\; \Ns_- = \ml 0\\ -ic \mr e^{-m|x|}
\end{equation}
where $c$ is a real operator. In condensed matter theory, such real fermion operators are called Majorana operators. The number of spinor components has been cut in half, and there is a zero-energy fermion localized in the vicinity of $x = 0$.
\\\\
Let $m_F$ be the coefficient of $\half \bar\Ns\!\!\Ns$ in the Lagrangian of Eq.~(\ref{eq:majorana lagrangian}). In a system without spatial boundaries, one typically assumes that the $m_F = +m$ phase and the $m_F = -m$ phase describe the same quantum state, because the sign can be compensated by a transformation $\Ns \to \g^5\Ns$. The existence of the Majorana mode at the kink core means that, for a system with spatial boundaries, these phases are different: at the interface between the two states, there is an additional degree of freedom \cite{jackiw rebbi}. The sign of the fermion mass term will play a crucial role throughout this paper.
\\\\
Now imagine a scalar field profile of the following form:
\begin{equation}
g\,\phi(x) = \left\{\begin{matrix} -m\;,\;\;x > L\\ 0\;,\;\; x = L\\  +m\;,\;\; -L < x < L\\ 0\;,\;\; x = -L\\ -m\;,\;\; x < -L \end{matrix}\right.\;.
\end{equation}
At $x = L$ there is a real fermion, $c$, and at $x = -L$ there is another real fermion, $c'$. These can be paired up into a complex fermion annihilation operator,
\begin{equation}
f = c+ic'\;.
\end{equation}
If $|0\ket$ is the vacuum with energy $E_0$, then the state $f^\da |0\ket$ has an energy\footnote{This is because the two ends at $x = \pm L$ have to talk to each other in order to form a term in the Hamiltonian of the form $icc'$. The interior is fully gapped and admits only local interactions, so the amplitude for the two ends to interact is exponentially suppressed for $L \gg m^{-1}$.} $E_1-E_0 \sim e^{-mL}$.  In the limit $L \to \infty$ (namely, the thermodynamic limit), the state $f^\da|0\ket$ becomes degenerate with the vacuum. So if one thinks of these states as belonging to the boundaries of the $m_F > 0$ phase while considering the $m_F < 0$ phase as the ``ordinary" gapped phase, then this profile models a topologically nontrivial $1d$ system of length $2L$.
\\\\
But this 1$d$ system of length $2L$ is not yet an SPT state, because the gaplessness of the excitation is protected by the thermodynamic limit, not by the imposition of a global symmetry. To emphasize this point, consider two flavors of the above setup, indexed by a label $a = 1,2$. Then it is possible to write the local interactions $ic_1c_2$ and $ic_1'c_2'$ at $x = +L$ and $x = -L$, respectively. All excitations above the ground state are gapped, and this is a trivial phase (in the sense described earlier).
\\\\
For this two flavor system, impose a flavor-independent \textit{antiunitary} discrete symmetry, which may as well be called a peculiar version of time reversal that squares to $+1$:
\begin{equation}\label{eq:Z2T}
\zb_2^T:\qquad \Ns_a(t,x) \to \g^0\Ns_a(-t,x)\;,\;\; i \to -i\;.
\end{equation}
This transformation leaves $c_1c_2$ and $c_1'c_2'$ unchanged, but it flips the sign of the prefactor $i$ (whose presence in the Hamiltonian is required for hermiticity). Hence if this $\zb_2^T$ is imposed on the Lagrangian, then all fermion bilinears at the $x = \pm L$ boundaries will be forbidden. 
\\\\
This is true for an arbitrary number of flavors, $n\in \zb$. Each value of $n$ defines a distinct phase. So this setup describes a $1d$ SPT phase which is classified by an integer that labels the number of gapless edge modes.\footnote{The reader may want to verify that this setup satisfies the three conditions described earlier. Conditions (1) and (2) are obviously fulfilled. Condition (3) is fulfilled because the action of time reversal as in Eq.~(\ref{eq:Z2T}) cannot be implemented self-consistently on an independent $0d$ quantum system. Define the annihilation operator for a fermionic oscillator at $x = +L$ by $a \equiv c_1+ic_2$. The $\zb_2^T$ transformation flips the sign of $i$ but leaves $c_1$ \textit{and} $c_2$ invariant. Therefore, $\zb_2^T:\, a \to a^\da$, and time reversal does not commute with $(-1)^F$ when acting on physical states.} It is crucial to observe that the flavor-diagonal transformation in Eq.~(\ref{eq:Z2T}) leaves the bilinear $\bar\Ns_a\Ns_b$ invariant, so the bulk remains gapped.  
\\\\
The possible free-fermion SPT phases in various dimensions and with various global symmetries have already been enumerated \cite{classification, periodic table}. The question that connects this to the particle physics problem described earlier is whether those systems with integer classification are stable to interactions. In condensed matter physics one is typically concerned only with time reversal, $SU(2)$ spin symmetry, and particle-hole symmetry (or its incarnation as an artificial redundancy in superconducting theories). But if the transition between the trivial superconducting phase and the SPT phase is continuous, then it admits a field theoretic description, and the results obtained in that description hold for any system described by the same low-energy effective Lagrangian. 
\\\\
Just as the electron of the SM is protected by $SU(2)_{\text{weak}}\x U(1)_{\text{hypercharge}}$, here in the $1d$ Kitaev chain the $0d$ edge fermions are protected by $\zb_2^T$. By turning on local interactions for a system with $n$ flavors, is it possible to gap out these symmetry-protected edge modes \textit{without} breaking $\zb_2^T$ spontaneously? Kitaev and Fidkowski (KF) \cite{KF} showed that the answer is yes, if and only if $n = 8k$, $k\in \zb$. (The reader who is unfamiliar with this result should not worry: it will be discussed thoroughly in the body of this paper.) One says that the interactions ``reduce the classification" from $\zb$ to $\zb_8$.
\\\\
The purpose of this paper is to explore in greater detail the $``m = 0"$ manifold of the KF model purely within the continuum field theory description, with an eye toward extracting general lessons for interacting field theories in higher dimensions. Just as 1+1 interacting systems have proved insightful for studying confinement in higher dimensions, I hope that a thorough analysis in $1d$ will provide guidance for interacting fermions in 3+1 dimensions.
\\\\
The layout is as follows. First, in Sec.~\ref{sec:GN}, I will review the $SO(8)$ Gross-Neveu model (GN). The purpose of this is to provide necessary background material, to establish notation, and to point out a subtlety in the ``triality" invariance of the Lagrangian. Then, in Sec.~\ref{sec:KF}, I will discuss the $``m = 0"$ manifold of the $SO(7)$ KF model with an emphasis on the fermion propagator. In particular, I will argue that an analog of ``parity doubling" occurs, and that the leading term in the spectral decomposition is simply proportional to $p^{\,\mu}$. In Sec.~\ref{sec:imp}, I will attempt to relate the KF model to physical conduction electrons in the context of impurity scattering. In Sec.~\ref{sec:end}, I will summarize the results and suggest possible directions for future work.  
\section{Eight Majorana fermions with $SO(8)$ symmetry}\label{sec:GN}
The goal is to study the effects of interactions on the mass gap and excitation spectrum for a theory of eight relativistic Majorana fermions, 
\begin{equation}
\Ns_a = \ml \bar\eta_a\\ i\eta_a \mr\;;\;\; a = 1,...,8\;.
\end{equation}
The free massless Lagrangian is:
\begin{equation}\label{eq:8 free massless}
\la_0 =\sum_{a\,=\,1}^8\half \bar\Ns_a i\!\!\not\!\pa \Ns_a = \sum_{a\,=\,1}^8 \half i \left( \eta_a\pa_-\eta_a+\bar\eta_a\pa_+\bar\eta_a\right)
\end{equation}
where $\pa_\pm \equiv \pa_t\pm\pa_x$. This Lagrangian has a continuous global symmetry $SO(8)_L\x SO(8)_R$. To the free Lagrangian in Eq.~(\ref{eq:8 free massless}), first add the following interaction, which breaks $SO(8)_L\x SO(8)_R$ down to the diagonal $SO(8)$:
\begin{equation}\label{eq:GN}
\la_{\text{int}}^{\text{(GN)}} = +\fourth g\left( \sum_{a\,=\,1}^8\bar\Ns_a\Ns_a\right)^2 = -g\left( \sum_{a\,=\,1}^8\eta_a\bar\eta_a\right)^2\;.
\end{equation}
From now on, the standard repeated index summation convention will be used.
\\\\
The Lagrangian $\la_{\text{GN}} = \la_0+\la_{\text{int}}^{\text{(GN)}}$ defines the $SO(8)$ Gross-Neveu (GN) model \cite{GN}. It has a global chiral $\zb_2$ symmetry:
\begin{equation}\label{eq:chiral Z2}
\zb_2:\qquad (\eta_a,\bar\eta_b) \to (-\eta_a,+\bar\eta_b)\;.
\end{equation}
However, this symmetry is spontaneously broken at low energy: the coupling gets strong and the fermion mass bilinear forms an $SO(8)$-invariant condensate,
\begin{equation}
\bra i\eta_a \bar\eta_b\ket = \pm v\,\del_{ab}\;,\;\; v > 0\;.
\end{equation}
Perturbing around a fixed choice of minimum,
\begin{equation}
i\eta_a\bar\eta_b = \pm v\,\del_{ab}\,+\, i\eta_a'\bar\eta_b'\;,
\end{equation}
one finds nonzero fermion masses for the fluctuations described by the primed fields:
\begin{equation}\label{eq:GN mass}
\la_{\text{int}}^{\text{(GN)}} = \text{const} - (\pm16g v) \sum_{a\,=\,1}^8 i\eta_a'\bar\eta_a'+...\;.
\end{equation}
This part of the story is the well-known analysis of the $SO(N)$ GN model at large $N$ (see, for example, \cite{GN large N}) and is not unique to the value $N = 8$.
\subsection{Bosonization and triality}\label{sec:triality}
The ``triality" of $SO(8)$ is a cyclic permutation of the three real 8-dimensional representations, which are the vector (denoted by $8_v$) and the two chiral spinors (denoted by $8_+$ and $8_-$) \cite{cartan}. Although the $SO(8)$ group possesses this outer automorphism, the physics of the $SO(8)$ GN model is a little more subtle. This is the same subtlety which occurs in the Ising model: a theory with two ground states cannot be equivalent to a theory with one ground state, so the Ising duality transformation must be accompanied by the introduction of a topological $\zb_2$ gauge theory \cite{Z2 spurion, coupling QFT to TQFT}. This will be reviewed in Sec.~\ref{sec:ising}, but first let me proceed with the $SO(8)$ theory.
\\\\
A physically clear way to implement the triality operation is to use abelian bosonization \cite{shankar so(8), maldacena so(8), balents so(8), ludwig so(8)}. (For a discussion of triality in non-abelian bosonization, see \cite{nepomechie so(8)}.) First bosonize the Majorana fermions in pairs:
\begin{equation}\label{eq:eta bosonization}
\eta_{2A-1}+i\eta_{2A} \equiv e^{\,i2\pi \ph_A}\;,\;\; \bar\eta_{2A-1}+i\bar\eta_{2A} \equiv e^{\,i2\pi \bar\ph_A}\;\;;\;\; A = 1,...,4\;.
\end{equation}
The chiral bosons $\ph_A(x+t)$ and $\bar\ph_A(x-t)$ are \textit{defined} by the above relations. So if the original physical model is given by Eq.~(\ref{eq:GN}), then the bosons are compact and are defined only modulo shifts by integers.\footnote{In Eq.~(\ref{eq:eta bosonization}) a non-standard normalization for the bosons has been chosen, because the additional factor of $\pi^{1/2}$ would needlessly clutter the discussion.}
\\\\
Define the non-chiral bosons
\begin{equation}
\Phi_A(x,t) \equiv \bar\ph_A(x+t)-\ph_A(x-t)\;.
\end{equation}
Then the $SO(8)$-invariant fermion mass term is:
\begin{equation}
 \sum_{a\,=\,1}^8i\eta_a\bar\eta_a = \sum_{A\,=\,1}^4 \cos(2\pi\Phi_A)\;.
\end{equation}
The triality transformation from the $8$-vector to the $8_+$-spinor is defined by the following special orthogonal transformation in the space of bosons (I use the conventions of Ludwig and Maldacena \cite{maldacena so(8)}):
\begin{equation}\label{eq:8+ matrix}
\ml \Phi_1\\ \Phi_2\\ \Phi_3\\ \Phi_4 \mr \equiv \frac{1}{2}\ml +1&+1&+1&+1\\ +1&-1&+1&-1\\ +1&+1&-1&-1\\ +1&-1&-1&+1 \mr \ml \Ta_1\\ \Ta_2\\ \Ta_3\\ \Ta_4 \mr\;.
\end{equation}
This change of basis \textit{defines} the bosons $\Ta_I$, $I = 1,...,4$. (To dispel any potential confusion, I should note that I will not use the notation ``$\Ta$" for the dual of $\Phi$. For the dual of $\Phi$ I will write $\tilde\Phi \equiv \ph_A+\bar\ph_A$.)
\\\\
By straightforward algebra, one obtains:
\begin{equation}
\sum_{A\,=\,1}^4 \cos(2\pi\Phi_A) = 4\left(  \prod_{I\,=\,1}^4 \cos(\pi\Ta_I)+\prod_{I\,=\,1}^4\sin(\pi\Ta_I)\right)\;.
\end{equation}
Therefore:
\begin{equation}
\left( \sum_{A\,=\,1}^4 \cos(2\pi\Phi_A)\right)^2_{\text{REN}} = \;\;\left( \sum_{I\,=\,1}^4 \cos(2\pi\Ta_I)\right)^2_{\text{REN}}\;.
\end{equation}
By the subscript ``REN" I mean that this equality holds after the renormalization procedure of subtracting the cosine-squared terms from both sides. The reason for doing this is because the quantum theory possesses the unusual relation (see the appendix of \cite{properties of 4-fermion}):
\begin{equation}
\cos^2(2\pi\Phi_A) \propto -\half (\pa_\mu\Phi_A)^2+\text{constant}\;.
\end{equation}
So these terms actually contribute to a renormalization of the boson kinetic terms and should not be considered as part of the interactions. 
\\\\
In analogy with the definition $\Phi_A = \ph_A-\bar\ph_A$, now define the chiral bosons $\ta_I$ and $\bar\ta_I$ via $\Ta_I \equiv \ta_I-\bar\ta_I$ and $\tilde\Ta_I \equiv \ta_I+\bar\ta_I$. These new chiral bosons can be fermionized: \footnote{A spinor of $SO(2n)$ should pick up a minus sign after a rotation through $2\pi$ in the $2n$-dimensional Euclidean embedding space. Consider a rotation by $2\pi$ in the $(1,2)$-plane. This corresponds to a shift $\Phi_1 \to \Phi_1+1$ with $\Phi_{2,3,4}$ fixed. From the inverse of Eq.~(\ref{eq:8+ matrix}) one finds $\Ta_I \to \Ta_I+\half$ for all $I = 1,2,3,4$. So the $8_+$ fermions in Eq.~(\ref{eq:8+ fermions}) indeed pick up a factor of $(-1)$. The same is true for the $8_-$ fermions in Eq.~(\ref{eq:8- fermions}).}
\begin{equation}\label{eq:8+ fermions}
e^{\,i2\pi \ta_I} \equiv \psi_{2I-1}+i\psi_{2I}\;,\;\;e^{\,i2\pi \bar\ta_I} \equiv \bar\psi_{2I-1}+i\bar\psi_{2I}\;\;;\;\; I = 1,...,4\;.
\end{equation}
So after the subtraction described above, the following equality is obtained:
\begin{equation}
\left( \sum_{a\,=\,1}^8 \eta_a\bar\eta_a\right)^2 = \left( \sum_{i\,=\,1}^8 \psi_i\bar\psi_i\right)^2\;.
\end{equation}
This process can be repeated starting from a modified version of Eq.~(\ref{eq:8+ matrix}):
\begin{equation}\label{eq:8- matrix}
\ml \Phi_1\\ \Phi_2\\ \Phi_3\\ \Phi_4 \mr \equiv \frac{1}{2}\ml +1&+1&+1&-1\\ +1&+1&-1&+1\\ +1&-1&+1&+1\\ -1&+1&+1&+1 \mr \ml \Xi_1\\ \Xi_2\\ \Xi_3\\ \Xi_4 \mr\;.
\end{equation}
Then:
\begin{equation}
\sum_{A\,=\,1}^4 \cos(2\pi\Phi_A) = 4\left(  \prod_{X\,=\,1}^4 \cos(\pi\Xi_X)-\prod_{X\,=\,1}^4\sin(\pi\Xi_X)\right)
\end{equation}
and 
\begin{equation}
\left( \sum_{A\,=\,1}^4 \cos(2\pi\Phi_A)\right)^2_{\text{REN}} = \;\;\left( \sum_{X\,=\,1}^4 \cos(2\pi\Xi_X)\right)^2_{\text{REN}}\;.
\end{equation}
Again it is useful to define chiral bosons via $\Xi_X = \xi_X-\bar\xi_X$ and fermionize them:
\begin{equation}\label{eq:8- fermions}
e^{\,i2\pi \xi_X} \equiv \chi_{2X-1}+i\chi_{2X}\;,\;\;e^{\,i2\pi \bar\xi_X} \equiv \bar\chi_{2X-1}+i\bar\chi_{2X}\;\;;\;\;X = 1,...,4\;.
\end{equation}
Therefore \cite{shankar so(8), shankar triality, complete S-matrix}:
\begin{equation}\label{eq:polynomial triality}
\left( \sum_{a\,=\,1}^8 \eta_a\bar\eta_a\right)^2 = \left( \sum_{i\,=\,1}^8 \psi_i\bar\psi_i\right)^2 = \left(\sum_{x\,=\,1}^8\chi_x\bar\chi_x\right)^2\;.
\end{equation}
The kinetic terms also satisfy an analogous equality, so the whole Lagrangian takes the same form whether written in terms of the $\eta$, the $\psi$, or the $\chi$ fermions. These fields are nonlocally related to each other, but the Lagrangian written in terms of a given representation is local.
\\\\
This is what is usually considered the physical manifestation of the group-theoretic triality symmetry of the $SO(8)$ GN model. The equality of the fourth-order polynomials in Eq.~(\ref{eq:polynomial triality}) was just derived explicitly above, so this part of the usual story remains unchallenged. I simply wish to point out a subtlety in the analysis if one studies the system in terms of the $\psi$-variables: the discrete ``$\g^5$" transformation $\psi\bar\psi \to-\psi\bar\psi$ is actually a gauge symmetry.
\subsection{Global $\zb_2$ symmetry and emergent $\zb_2'$ gauge symmetry}\label{sec:Z2}
Consider the global chiral $\zb_2$ symmetry defined back in Eq.~(\ref{eq:chiral Z2}). This corresponds to a shift 
\begin{equation}\label{eq:chiral Z2 bosons}
\zb_2:\qquad (\ph_A,\bar\ph_A) \to (\ph_A+\half,\bar\ph_A)
\end{equation}
for all $A = 1,2,3,4$ simultaneously. The goal is to determine how this transformation affects the fields $\psi \sim 8_+$ and $\chi \sim 8_-$.
\\\\
Recall the transformations in Eqs.~(\ref{eq:8+ matrix}) and~(\ref{eq:8- matrix}), which I repeat below for convenience:
\begin{align}
&\vec\ph = S\, \vec\ta = T\, \vec\xi\;,\;\;S = \frac{1}{2}\ml +1&+1&+1&+1\\ +1&-1&+1&-1\\ +1&+1&-1&-1\\ +1&-1&-1&+1 \mr\;,\;\; T = \frac{1}{2}\ml +1&+1&+1&-1\\ +1&+1&-1&+1\\ +1&-1&+1&+1\\ -1&+1&+1&+1 \mr\;. \label{eq:triality matrices}
\end{align}
I have written this relation in terms of the left-moving chiral bosons, because it is these which are shifted by the chiral $\zb_2$ transformation. The matrices $S$ and $T$ satisfy $S^2 = I$ and $T^2 = I$, and hence $S = S^{-1}$, $T = T^{-1}$. Therefore, in terms of the original chiral bosons $\{\ph_A\}_{A\,=\,1}^4$, the defining relations above imply:
\begin{equation}
\ml \ta_1\\ \ta_2\\ \ta_3\\ \ta_4 \mr = \frac{1}{2}\ml \ph_1+\ph_2+\ph_3+\ph_4\\ \ph_1-\ph_2+\ph_3-\ph_4\\ \ph_1+\ph_2-\ph_3-\ph_4\\ \ph_1-\ph_2-\ph_3+\ph_4 \mr\;,\;\;\ml \xi_1\\ \xi_2\\ \xi_3\\ \xi_4 \mr = \frac{1}{2}\ml \ph_1+\ph_2+\ph_3-\ph_4\\ \ph_1+\ph_2-\ph_3+\ph_4\\ \ph_1-\ph_2+\ph_3+\ph_4\\ -\ph_1+\ph_2+\ph_3+\ph_4 \mr\;.
\end{equation}
If $\ph_A \to \ph_A + \half$, then
\begin{equation}\label{eq:chiral Z2 other bosons}
\zb_2:\qquad \ta_I \to \ta_I\;\;(\text{mod } 1)\;,\;\; \xi_X \to \xi_X +\half\;.
\end{equation}
Upon refermionization as in Eqs.~(\ref{eq:8+ fermions}) and~(\ref{eq:8- fermions}), I conclude that the physical $\zb_2$ symmetry acts as follows on the $8_\pm$ fermions:
\begin{equation}
\zb_2:\qquad (\psi_i,\bar\psi_j) \to (+\psi_i,\bar\psi_j)\;,\;\; (\chi_x,\bar\chi_y) \to (-\chi_x,\bar\chi_y)\;.
\end{equation}
Therefore, the $8_+$ mass bilinear $\psi_i\bar\psi_j$ is \textit{even} and hence is not an order parameter for the $\zb_2$ symmetry. It may self-consistently obtain an expectation value without spontaneously breaking $\zb_2$. 
\\\\
On the other hand, the $8_-$ mass bilinear $\chi_x\bar\chi_y$ is odd and hence cannot obtain an expectation value if the $\zb_2$ transformation is to remain a symmetry of the low-energy theory. In this way, perhaps counterintuitively, the two $SO(8)$ spinors are not created equal: it is not possible to use the $8_-$ in order to connect the trivial and topological phases. 
\\\\
The point about the triality transformation is to consider the analogous chiral sign flip for the $8_+$ variables, which I will denote by $\zb_2'$. This operation is defined as
\begin{equation}\label{eq:chiral Z2'}
\zb_2':\qquad (\psi_i,\bar\psi_j) \to (-\psi_i, \bar\psi_j)\;.
\end{equation}
The goal is now to determine how this transforms the fields $\eta \sim 8_v$ and $\chi \sim 8_-$. To do this, it is necessary to express the chiral bosons $\ph_A$ and $\xi_X$ in terms of the $8_+$ bosons $\ta_I$, which transform as
\begin{equation}
\zb_2':\qquad \ta_I \to \ta_I+\half\;. 
\end{equation}
 Recalling the triality transformation in Eq.~(\ref{eq:triality matrices}), one finds (also recall that $T^{-1} = T$):
\begin{equation}
\vec\ph = S\,\vec\ta\;,\;\; \vec\xi = TS\,\vec\ta\;.
\end{equation}
The product of the two transformation matrices,
\begin{equation}
TS = \frac{1}{2}\ml +1&+1&+1&-1\\ +1&-1&+1&+1\\ +1&+1&-1&+1\\ +1&-1&-1&-1 \mr\;,
\end{equation}
contains an \textit{odd} number of minus signs per row. Meanwhile, the matrix $S$ has an \textit{even} number of minus signs per row. Therefore, the operation in Eq.~(\ref{eq:chiral Z2'}) shifts the other two sets of bosons as
\begin{equation}
\zb_2':\qquad \ml \ph_1\\ \ph_2\\ \ph_3\\ \ph_4 \mr \to \ml \ph_1\\ \ph_2\\ \ph_3\\ \ph_4 \mr+\ml 1\\ 0\\ 0\\0 \mr\;,\;\; \ml \xi_1\\ \xi_2\\ \xi_3\\ \xi_4 \mr \to \ml \xi_1\\ \xi_2\\ \xi_3\\ \xi_4 \mr+\frac{1}{2}\ml +1\\ +1\\ +1\\ -1 \mr\;.
\end{equation}
Since $e^{\,i\pi} = e^{-i\pi} = -1$, the relative sign in the transformation for the $\xi_X$ is immaterial, and I conclude:
\begin{equation}
\zb_2':\qquad \eta_a \to +\eta_a\;,\;\; \chi_x \to -\chi_x\;.
\end{equation}
Therefore, the alternative chiral reflection defined by Eq.~(\ref{eq:chiral Z2'}) leaves the original fermion fields $\eta_a$ totally unaffected. This transformation is invisible in terms of the original fields in the Lagrangian and hence should be thought of as an \textit{emergent gauge symmetry}.
\\\\
Finally, one should consider the theory written in terms of the $8_-$ variables and define a third $\zb_2$ transformation which acts as $(\chi_x,\bar\chi_y) \to (-\chi_x,\bar\chi_y)$. I will not give this operation a new name because it turns out to be equivalent to the original global $\zb_2$ symmetry. The by-now-familiar triality transformations give:
\begin{equation}
\vec\ph = T\,\vec \xi\;,\;\; \vec\ta = ST\,\vec\xi\;.
\end{equation}
The product of $S$ and $T$ in this order contains an \textit{even} number of minus signs per row,
\begin{equation}
ST = \frac{1}{2}\ml +1&+1&+1&+1\\ +1&-1&+1&-1\\ +1&+1&-1&-1\\ -1&+1&+1&-1 \mr\;,
\end{equation}
while the matrix $T$ contains an odd number of minus signs per row. Therefore, the shift $\xi_X \to \xi_X + \half$ results in the shifts
\begin{equation}
\ph_A\to \ph_A+\half\;,\;\; \ta_I \to \ta_I\;\;(\text{mod } 1)\;.
\end{equation}
This is exactly the same transformation as the one described by Eqs.~(\ref{eq:chiral Z2 bosons}) and~(\ref{eq:chiral Z2 other bosons}).
\\\\
It is convenient to summarize this situation in terms of the mass bilinears. There are two $\zb_2$ transformations, one global and one gauged. The physical $\zb_2$ global symmetry acts as
\begin{align}\label{eq:physical Z2}
\zb_2:\qquad & \eta_a\bar\eta_b \to -\eta_a\bar\eta_b \implies \left\{  \begin{matrix} \psi_i\bar\psi_j \to +\psi_i\bar\psi_j \\ \chi_x\bar\chi_y \to -\chi_x\bar\chi_y \end{matrix}\right.\;.
\end{align}
The artificial $\zb_2'$ gauge symmetry acts as
\begin{align}\label{eq:gauge Z2}
\zb_2':\qquad & \psi_i\bar\psi_j \to -\psi_i\bar\psi_j \implies \left\{ \begin{matrix} \eta_a\bar\eta_b \to +\eta_a\bar\eta_b \\ \chi_x\bar\chi_y \to -\chi_x\bar\chi_y  \end{matrix} \right.\;.
\end{align}
So, strictly speaking, the physics of the $SO(8)$ Gross-Neveu model is not quite invariant under triality: the description in terms of the $8_+$ variables requires coupling to a topological $\zb_2'$ gauge theory. This gauging procedure does not add any additional local degrees of freedom, but it projects out sectors of the state space which are not invariant under the transformation in Eq.~(\ref{eq:gauge Z2}). 
\subsection{Fermion parity}\label{sec:fermion parity}
In addition to the chiral $\zb_2$ transformation $\Ns_a \to \g^5\Ns_a$, it is also interesting to consider the transformation $\Ns_a \to -\g^5\Ns_a$. The analysis goes through exactly as before, except with the barred chiral fields playing the role of the unbarred chiral fields. The product of both of these transformations is fermion parity,
\begin{equation}
(-1)^F:\qquad \Ns_a \to -\Ns_a\;.
\end{equation} 
Therefore, the conclusions of the previous section imply that the fields $\Psi_i = \ml \bar\psi_i\\ i\psi_i \mr$ are \textit{even} under fermion parity, while the fields $\mathscr X_x = \ml \bar\chi_x\\ i\chi_x \mr$ are odd. 
\\\\
This presents a puzzle: if one wishes to describe the original theory of $\eta$ variables in terms of $\psi$ variables, how is it possible to recover the sector of the original Hilbert space which contains an odd number of fermions? It is clear that additional non-local data is required, and I do not yet have a complete solution to this problem.
\\\\
Furthermore, it is also interesting to consider the ``artificial" fermion parity,
\begin{equation}
(-1)^{F'}:\qquad \Psi_i \to -\Psi_i\;.
\end{equation}
This too is a gauge symmetry and should be modded out in the $\psi$-description of the original theory. 
\subsection{$\zb_2$ transformations in the Ising model}\label{sec:ising}
It is useful to recall various properties of the $2d$ Ising model \cite{kogut, itzykson, boyanovsky}. (The second ``spatial" direction in this context should be thought of as Euclidean time.) I will work in the extreme anisotropic limit, which admits a description in terms of a transfer matrix (formally equivalent to deriving the path integral formulation by cutting up the total time interval into a large number of arbitrarily small steps).
\\\\
In the transfer matrix description, there is a $1d$ lattice labeled by sites 
\begin{equation}
s \in\{ 1,...,N\}\;,\;\; N\gg 1\;.
\end{equation}
On each site lives a ``spin" variable $\s_s$ which can be up or down, denoted by $+1$ and $-1$ respectively. In operator language, I choose this to be an eigenvalue of the third Pauli operator, $\hat\s_s^z$. The states $|\prod_{s\,=\,1}^N \s_s\ket \equiv |\s_1\ket\ox|\s_2\ket\ox...\ox|\s_N\ket$ satisfy:
\begin{equation}
\hat\s_s^z |\s_1\s_2...\s_N\ket = \s_s|\s_1\s_2...\s_N\ket\;.
\end{equation}
The Hamiltonian is:
\begin{equation}\label{eq:ising model}
\hat H = -\sum_{s\,=\,1}^{N-1} \hat\s_s^z\hat\s_{s+1}^z-\ld\sum_{s\,=\,1}^N \hat\s_s^x\;.
\end{equation}
Free boundary conditions have been chosen in the spatial direction. The low temperature phase is described by $\ld \ll 1$, and the high temperature phase is described by $\ld \gg 1$. The critical point is $\ld = 1$.
\\\\
The Hamiltonian has the following reflection symmetry:
\begin{equation}\label{eq:Z2spin}
\zb_2^{\text{spin}}:\qquad \hat\s_s^z \to -\hat\s_s^z
\end{equation}
for all $s = 1,...,N$ simultaneously. There are two possible ground states: all spins are aligned, and they all point either up or down:
\begin{equation}
|0\ket_\uparrow \equiv |++\;...\;+\ket\;,\;\; |0\ket_\downarrow \equiv |--\;...\;-\ket\;.
\end{equation}
The transformation of Eq.~(\ref{eq:Z2spin}) exchanges these states:
\begin{equation}
\zb_2^{\text{spin}}:\qquad |0\ket_\uparrow \leftrightarrow |0\ket_\downarrow\;.
\end{equation}
The total spin operator, or ``magnetization" (normalized by the number of sites),
\begin{equation}
\hat M \equiv \frac{1}{N}\sum_{s\,=\,1}^N\hat\s_s^z\;,
\end{equation}
has a nonzero vacuum expectation value:
\begin{equation}
_\uparrow\bra 0|\hat M|0\ket_\uparrow = +1\;,\;\; _\downarrow\bra 0|\hat M|0\ket_\downarrow = -1\;.
\end{equation}
In either case, the system is ordered (or ``magnetized"). The global symmetry $\zb_2^{\text{spin}}$ is broken spontaneously.
\\\\
A local excitation above one of the two ground states is given by the flip of a single spin. One can also consider a \textit{non}-local type of excitation, in which all spins to the left of a specified site, say $r$, are flipped. This excitation is called a kink (or domain wall), and is formally created by the following operator:
\begin{equation}\label{eq:mu z}
\hat\mu_{\tilde r}^z \equiv \prod_{s\,=\,1}^{r}\hat\s_s^x\;.
\end{equation}
These ``dual spins" live between the sites on the original lattice, which defines a dual lattice with $N-1$ sites:
\begin{equation}
\tilde r\in\{1,...,N-1\}\;.
\end{equation}
If one also defines the operator
\begin{equation}\label{eq:mu x}
\hat\mu_{\tilde r}^x \equiv \hat\s_r^z \hat\s_{r+1}^z
\end{equation}
then the $\mu$ variables define a good collection of Pauli matrices, and the Hamiltonian becomes:
\begin{equation}\label{eq:dual ising model}
\hat H = \ld\left(-\sum_{\tilde r\,=\,1}^{N-2} \hat\mu_{\tilde r}^z\hat\mu_{\tilde r+1}^z-\frac{1}{\ld}\sum_{\tilde r\,=\,1}^{N-1} \hat\mu_{\tilde r}^x\right)-\ld(\hat\s_1^x+\hat\s_N^x)\;.
\end{equation}
Comparison of this with Eq.~(\ref{eq:ising model}) reveals that the bulk energy spectrum obeys $E(\ld) = \ld E(1/\ld)$, showing the equivalence between the high temperature and low temperature phases. This is well-known, but I wish to emphasize the following subtlety regarding this description in terms of the $\mu$ variables \cite{Z2 spurion, coupling QFT to TQFT}. 
\\\\
In analogy with the original description, define states in terms of ``dual" spins, meaning eigenvalues of $\hat \mu_r^z$:
\begin{equation}
\hat \mu_{\tilde r}^z|\tilde \s_1\tilde \s_2...\tilde\s_N\ket = \tilde \s_{\tilde r}|\tilde \s_1\tilde \s_2...\tilde\s_N\ket\;,\;\; \tilde\s_{\tilde r}\in\{-1,+1\}\;.
\end{equation}
At $1/\ld = 0$, there appear to be two possible ground states:
\begin{equation}
|\tilde 0\ket_\uparrow \equiv |\tilde +\tilde +\;...\; \tilde +\ket\;,\;\; |\tilde 0\ket_\downarrow \equiv |\tilde -\tilde -\;...\;\tilde -\ket\;.
\end{equation}
But the transformation $\ld \leftrightarrow 1/\ld$ exchanges the high and low temperature phases, and the high temperature (disordered) phase of the original Ising model is \textit{unique}. Therefore, the dual of Eq.~(\ref{eq:Z2spin}), namely the transformation
\begin{equation}\label{eq:Z2dualspin}
\tilde\zb_2^{\text{spin}}:\qquad \hat\mu_{\tilde r}^z \to -\hat\mu_{\tilde r}^z\qquad\text{for all $\tilde r = 1,...,N-1$ simultaneously}
\end{equation}
must be gauged. This can be seen by explicitly calculating the operator which flips all of the dual spins. Using the definition in Eq.~(\ref{eq:mu x}), one has:
\begin{equation}\label{eq:Q}
\hat Q \equiv \hat \mu_1^x\;\hat \mu_2^x\; ... \;\hat \mu_{N-1}^x = \hat \s_1^x \; 1_2 1_3\,...\, 1_{N-1}\;\hat \s_N^x\;.
\end{equation}
I have written the factors of $1_{\tilde r}$ to emphasize that, by direct computation, one observes that the operation of flipping all dual spins simultaneously is simply the identity operator in the bulk.
\\\\
The situation is summarized as follows. If one begins with Eq.~(\ref{eq:ising model}), then this simply describes an Ising model (by definition). If one begins with Eq.~(\ref{eq:dual ising model}), then this also simply describes an Ising model, with a trivial change of Greek letters from $\s$ to $\mu$. However, if one wishes to describe the partition function corresponding to Eq.~(\ref{eq:ising model}) using the dual Hamiltonian in Eq.~(\ref{eq:dual ising model}), then one must also impose the operator relations Eqs.~(\ref{eq:mu z}) and~(\ref{eq:mu x}). 
\\\\
Equivalently, to describe the original Ising model in terms of the dual variables, it is necessary to impose a ``Gauss's law" constraint on the physical states, $|\Psi\ket_{\text{phys}}$:
\begin{equation}
\hat Q|\Psi\ket_{\text{phys}} = |\Psi\ket_{\text{phys}}\;,
\end{equation}
where $\hat Q$ is the operator defined in Eq.~(\ref{eq:Q}). This implies:
\begin{equation}
\hat Q|\tilde 0\ket_\uparrow = |\tilde 0\ket_\downarrow\;,\;\; \hat Q|\tilde 0\ket_\downarrow = |\tilde 0\ket_\uparrow\;.
\end{equation}
The physical ground state of the dual model is then the $\hat Q$-invariant superposition
\begin{equation}\label{eq:0phys}
|\tilde 0\ket_{\text{phys}} \equiv \sq\left( |\tilde 0\ket_\uparrow+|\tilde 0\ket_\downarrow\right)\;.
\end{equation}
The orthogonal combination,
\begin{equation}\label{eq:0unphys}
|\tilde 0\ket_{\text{unphys}} \equiv \sq\left( |\tilde 0\ket_\uparrow-|\tilde 0\ket_\downarrow\right)\;,
\end{equation}
is not gauge invariant and hence is projected out of the Hilbert space. The total dual spin, or ``disorder parameter," 
\begin{equation}
\hat K \equiv \frac{1}{N-1}\sum_{\tilde r\,=\,1}^{N-1} \hat \mu_{\tilde r}^z\;, 
\end{equation}
has nonzero expectation value in each of the two gauge-\textit{variant} states:
\begin{equation}\label{eq:kink condensation}
_\uparrow\bra \tilde 0|\hat K|\tilde 0\ket_\uparrow = +1\;,\;\; _\downarrow\bra\tilde 0|\hat K|\tilde 0\ket_\downarrow = -1\;.
\end{equation}
It is in this sense that the $\zb_2^{\text{spin}}$-symmetric disordered phase of the Ising model is recovered by the condensation of kinks. However, strictly speaking, the kink operator has zero expectation value in the physical ground state:
\begin{equation}
_{\text{phys}}\bra \tilde 0|\hat K|\tilde 0\ket_{\text{phys}} = 0\;.
\end{equation}
This is consistent with the general principle that gauge symmetries can never actually be broken spontaneously \cite{elitzur, wilczek CMT QCD}. But, as indicated for example by Eq.~(\ref{eq:kink condensation}), it is often extremely convenient to fix a gauge and to use the terminology which is more correctly reserved for the spontaneous breaking of global symmetries. 
\\\\
In summary, the Kramers-Wannier (KW) duality transformation
\begin{equation}\label{eq:KW}
\zb_2^{\text{KW}}:\qquad \hat \s_s^z \to \hat \mu_s^z
\end{equation}
must be accompanied by the introduction of a topological $\tilde\zb_2^{\text{spin}}$ gauge theory which implements the constraint given by the last equality in Eq.~(\ref{eq:Q}). At the critical point, $\ld = 1/\ld$, the bulk Hamiltonian is formally invariant under Eq.~(\ref{eq:KW}), but the global structure of the partition function must be modified to correctly reproduce Eq.~(\ref{eq:0phys}). 
\\\\
For the purpose of this paper, it is essential to recall the fermionic description of this model. The fermionization can be viewed as a solution to the duality algebra
\begin{equation}
\hat \mu_{\tilde r}^z\; \hat\s_s^z-(-1)^{\ta(\tilde r-s)}\;\hat \s_s^z\;\hat\mu_{\tilde r}^z = 0
\end{equation}
in terms of unconstrained variables \cite{Z2 spurion}. That the Hamiltonian written in terms of those fermion variables is local and quadratic constitutes the miracle of the Ising model \cite{itzykson}. 
\\\\
Define the operators
\begin{equation}
\hat \s_s^\pm \equiv \hat\s_s^z\pm i\hat\s_s^y\;.
\end{equation}
Then the operators
\begin{equation}\label{eq:ising fermions}
\hat f_s \equiv \left(\prod_{r\,=\,1}^{s-1} \hat \s_r^x\right) \half\hat \s_s^+ = \left(e^{-i\frac{\pi}{4}\sum_{r\,=\,1}^{s-1} \hat \s_r^+\hat \s_r^-}\right) \half\hat\s_s^+
\end{equation}
satisfy canonical anticommutation relations:
\begin{equation}
\{\hat f_s,\,\hat f_{s'}^\da\} = \del_{ss'}\;,\;\;\{\hat f_s,\,\hat f_{s'}\} = 0\;.
\end{equation}
By direct computation, one finds $(\hat f_s-\hat f_s^\da)(\hat f_{s+1}+\hat f_{s+1}^\da) = \hat \s_s^z\hat\s_{s+1}^z$ and $2\hat f_s^\da \hat f_s-1 = \hat\s_s^x$, and therefore the Ising Hamiltonian of Eq.~(\ref{eq:ising model}) can be expressed as a quadratic function of fermion operators:
\begin{equation}
\hat H = -\sum_{s\,=\,1}^{N-1}(\hat f_s-\hat f_s^\da)(\hat f_{s+1}+\hat f_{s+1}^\da)-\ld\sum_{s\,=\,1}^N(2 \hat f_s^\da \hat f_s-1)\;.
\end{equation}
Define the real and imaginary parts of the fermionic operators in Eq.~(\ref{eq:ising fermions}):
\begin{equation}
\hat f_s \equiv \hat c_{2s-1}+i \hat c_{2s}\;,\;\; \hat c_a^\da = \hat c_a\;,\;\; \{\hat c_a, \hat c_b\} = \half \del_{ab}\;,
\end{equation}
where $a,b\in\{1,...,2N\}$. Then the Hamiltonian takes the form of the Kitaev chain with $\ld = J_1/J_2$ (see Appendix~\ref{sec:kitaev chain}):
\begin{equation}
\hat H = -4\left( \sum_{s\,=\,1}^{N-1}i \hat c_{2s} \hat c_{2s+1}+\ld \sum_{s\,=\,1}^N i \hat c_{2s-1} \hat c_{2s}\right)\;.
\end{equation}
The continuum limit is described by the Lagrangian for a free Majorana fermion:
\begin{equation}\label{eq:free majorana lagrangian}
\la = \half \bar\Ns(i\!\!\not\!\pa-m)\Ns\;,\;\; m = \ld-1\;.
\end{equation}
The KW duality exchanges $\ld > 1$ with $\ld < 1$ and hence changes the sign of the fermion mass term:
\begin{equation}
\zb_2^{\text{KW}}:\qquad \bar\Ns\!\Ns \to -\bar\Ns\!\Ns\;.
\end{equation}
This is the ``$\g^5$" transformation that emerges when the fermion mass term is tuned to zero. 
\subsection{Ground state degeneracy in GN}\label{sec:2fold degeneracy}
In preparation for a later discussion of the $SO(7)$ Kitaev-Fidkowski model (Sec.~\ref{sec:degeneracy}),  it will be important to establish that the ground state of the $SO(8)$ GN model is two-fold degenerate. The potential written in terms of the bosons for the original Majorana fermion fields (the $\eta_a, \bar\eta_a$) is:
\begin{equation}\label{eq:GN potential}
V = -2g\sum_{A<B} \cos(2\pi\Phi_A)\cos(2\pi\Phi_B)\;.
\end{equation}
This potential is invariant under the simultaneous sign flip of all $\cos(2\pi\Phi_A)$. (This is just the physical $\zb_2$ symmetry that I have already discussed at length). The minima occur when all cosine terms equal $+1$ or $-1$. Following the terminology of Shankar \cite{shankar so(8)}, I will call these ``positive vacua" and ``negative vacua" respectively.
\\\\
Recall that the bosons $\Phi_A$ were defined by the relations in Eq.~(\ref{eq:eta bosonization}), so each $\Phi_A$ is defined only modulo 1. Therefore, all of the positive vacua correspond to a single state $|0\ket_>$ in the Hilbert space labeled by the configuration $\Phi_A = 0$ for all $A \in \{1,2,3,4\}$:
\begin{equation}\label{eq:positive vacuum}
|0\ket_> \qquad\leftrightarrow \qquad\Phi_A = (0,0,0,0)\;.
\end{equation}
For the same reason, all of the negative vacua also correspond to a single state $|0\ket_<$ labeled by the configuration $\Phi_A = \half$ for all $A \in \{1,2,3,4\}$:
\begin{equation}\label{eq:negative vacuum}
|0\ket_<\qquad \leftrightarrow\qquad \Phi_A = (\half,\half,\half,\half)\;.
\end{equation}
There are two degenerate ground states, corresponding to the spontaneous breaking of the global $\zb_2$ symmetry that interchanges them.
\subsection{Kinks and the $8_+$ basis}
In the $SO(8)$ GN model, the chiral $\zb_2$ symmetry is spontaneously broken by an $SO(8)$-invariant fermion condensate $\bra i\eta_a\bar\eta_b\ket = \pm v\, \del_{ab}$. There are 16 different kink configurations, which interpolate from $\eighth \sum_{a\,=\,1}^8i\eta_a\bar\eta_a = +v$ at $x = -\infty$ to $\eighth \sum_{a\,=\,1}^8i\eta_a\bar\eta_a = -v$ at $x = +\infty$ \cite{DHN kinks, mandelstam kinks, properties of 4-fermion, complete S-matrix}. These 16 kinks transform as $8_+\oplus 8_-$ under $SO(8)$ and are precisely the fermions $\psi_i$ and $\chi_x$.
\\\\
Triality suggests that it should be possible to arrive at the conclusion of the previous section by studying the GN model in terms of these kink fields. In the $8_+$ basis, the $SO(8)$ GN model has the Lagrangian
\begin{equation}
\la(\psi,\bar\psi) = \sum_{i\,=\,1}^8\half i(\psi_i\pa_-\psi_i+\bar\psi_i\pa_+\bar\psi_i)-g\left(\sum_{i\,=\,1}^8 \psi_i\bar\psi_i \right)^2\;.
\end{equation}
As discussed, it is understood that this should be coupled to a topological $\zb_2'$ gauge theory. Up to this subtlety, this Lagrangian looks formally equivalent to $\la(\eta,\bar\eta)$, so the local dynamics are the same: at low energy the theory forms an $SO(8)$-invariant condensate, 
\begin{equation}\label{eq:8+ condensate}
\bra i\psi_i\bar\psi_j\ket = \pm v\,\del_{ij}\;.
\end{equation}
Suppose the $\zb_2'$ transformation were not gauged. Then the two choices of sign in Eq.~(\ref{eq:8+ condensate}) would correspond to different ground states, just like the two minima of a standard double-well potential. Denote these two ground states by $|0\ket_+$ and $|0\ket_-$. The $\zb_2'$ transformation exchanges these two states:
\begin{equation}
\zb_2':\qquad |0\ket_{\pm} \to |0\ket_{\mp}\;.
\end{equation}
The physical implication of gauging the $\zb_2'$ symmetry is that the ground state is in fact the gauge-invariant linear superposition of the two possible configurations:
\begin{equation}
|0\ket_{\text{phys}} = \sq\left( |0\ket_+ +|0\ket_-\right)\;.
\end{equation}
The orthogonal combination, $|0\ket_{\text{unphys}} = \sq\left( |0\ket_+ -|0\ket_-\right)$, is not gauge invariant and hence is projected out of the Hilbert space. 
\\\\
But the conclusion of Sec.~\ref{sec:2fold degeneracy} was that the two possible choices of sign in the $\eta\bar\eta$ condensate do correspond to different physical ground states. How can one arrive at this conclusion from studying the $\psi$ variables? 
\\\\
For this purpose it is useful to think of the $\psi_1,...,\psi_8$ as eight Ising fermions \cite{ising fermions, zuber itzykson, schroer truong, polyakov CFT}. Then, from the bosonization rules, one finds \cite{boyanovsky}:
\begin{equation}\label{eq:order disorder}
\sum_{a\,=\,1}^8i\eta_a\bar\eta_a \propto \prod_{i\,=\,1}^8 \s_i^{(\psi)}+\prod_{i\,=\,1}^8\mu_i^{(\psi)} \;.
\end{equation}
In this model, the condensate in Eq.~(\ref{eq:8+ condensate}) induces an $SO(8)$-invariant mass for the $\psi_i$ variables. Just as in Eq.~(\ref{eq:GN mass}), one expands around the condensate,
\begin{equation}
\psi_i\psi_j = \bra \psi_i\psi_j\ket + \psi_i'\psi_j'\;,
\end{equation}
and finds a nonzero mass term for the fluctuations:
\begin{equation}
\la(\psi',\bar\psi') = \sum_{i\,=\,1}^8\half i(\psi_i'\pa_i\psi_i'+\bar\psi_i'\pa_+\bar\psi_i')-(\pm 16gv) \sum_{i\,=\,1}^8 i\psi_i'\bar\psi_i'+...\;.
\end{equation}
The mass parameter for an Ising fermion is proportional to $T-T_c$:
\begin{equation}
m_{\text{Ising}} \propto T-T_c\;.
\end{equation}
 For the ``$+$" sign, the corresponding Ising models are in their \textit{disordered} phase: $\bra \s_1^{(\psi)}\ket = \bra\s_2^{(\psi)}\ket = ... = \bra\s_8^{(\psi)}\ket = 0$, while $\bra \mu_1^{(\psi)}\ket = \bra\mu_2^{(\psi)}\ket = ... = \bra\mu_8^{(\psi)}\ket \neq 0$. For the ``$-$" sign, they are in their \textit{ordered} phase: $\bra \s_1^{(\psi)}\ket = \bra\s_2^{(\psi)}\ket = ... = \bra\s_8^{(\psi)}\ket \neq 0$, while $\bra \mu_1^{(\psi)}\ket = \bra\mu_2^{(\psi)}\ket = ... = \bra\mu_8^{(\psi)}\ket = 0$. Either way, the $SO(8)$ symmetry requires all eight Ising models to be in the same phase, either ordered or disordered, so at low energy one always has:
\begin{equation}
\left\bra\sum_{a\,=\,1}^8 \eta_a\bar\eta_a\right\ket \neq 0\;.
\end{equation}
The $\zb_2$ operation which flips the sign of this bilinear is a physical symmetry (not a gauge redundancy) and transforms a given ground state into another inequivalent ground state. This is one way to arrive at the conclusion of Sec.~\ref{sec:2fold degeneracy} from the $8_+$ basis. 
\\\\
For the $SO(8)$ GN model, this argument was needlessly complicated: one could have just analyzed the theory in terms of the original $\eta$ variables and arrived at the correct conclusion directly. The purpose of this exercise was to show that the formation of a condensate in $\psi\bar\psi$ does not necessarily imply ground state degeneracy. The ground state may or may not be unique, irrespective of whether $\bra\psi\bar\psi\ket = 0$. 
\section{Eight Majorana fermions with $SO(7)$ symmetry}\label{sec:KF}
After this long but necessary preliminary discussion of the $SO(8)$ Gross-Neveu model, I can now proceed to the $SO(7)$ Kitaev-Fidkowski model. 
\\\\
From the discussion surrounding Eq.~(\ref{eq:order disorder}), it is clear that the goal should be to single out a direction in the kink basis. Furthermore, the transformation properties under $\zb_2$ [Eq.~(\ref{eq:physical Z2})] indicate that only the $8_+$ can form a fermion bilinear condensate without generating a mass term for the $\eta$ variables at some order in perturbation theory. 
\\\\
Therefore, the appropriate course of action is to add an additional four-fermion interaction to the GN model which singles out a direction in the $8_+$ representation \cite{KF, KF 2}. The KF Lagrangian is $\la = \la_0+\la_{\text{int}}^{\text{(GN)}}+\la_{\text{int}}^{\text{(KF)}}$, where the additional interaction term is:
\begin{equation}\label{eq:KF}
\la_{\text{int}}^{\text{(KF)}} = -g'\!\!\!\!\!\sum_{a,b,c,d\,=\,1}^8\!\!\!\!\bra S |\G^{[a}\G^{b}\G^{c}\G^{d]}|S\ket\, \eta_a\bar\eta_b\eta_c\bar\eta_d\;.
\end{equation}
The symbols $\G^a$ denote the gamma matrices for $SO(8)$: there are 8 of these, and each is a matrix of size 16$\x$16. The state $|S\ket$ is defined as a particular element of the $8_+$; in the Wilczek-Zee notation \cite{wilczek zee, spinors}, the choice in this paper (and in \cite{KF}) is:
\begin{equation}
|S\ket = \sq \left( |++++\ket - |----\ket\right)\;.
\end{equation}
The basis of field coordinates $\psi_1,...,\psi_8$ is chosen so that $\psi_8$ corresponds to the state $|S\ket$. (In other words, I could self-consistently choose the notation $\psi_S \equiv \psi_8$.) The square brackets around the indices $a,b,c,d$ denote complete antisymmetrization.
\\\\
The interaction in Eq.~(\ref{eq:KF}) explicitly breaks the $SO(8)$ symmetry but conserves the $SO(7)$ subgroup which rotates among the 7 remaining states of the $8_+$ representation (namely those states which are orthogonal to $|S\ket$). Therefore, as explained in \cite{KF}, a triality transformation (understood in the sense discussed previously) to the $\psi$-fermion basis must result in a local polynomial of the form:
\begin{equation}
\la_{\text{int}}^{\text{(GN)}}+\la_{\text{int}}^{\text{(KF)}} = -\As\left( \sum_{i\,=\,1}^7 \psi_i\bar\psi_i\right)^2\!\!-\Bs \left( \sum_{i\,=\,1}^7 \psi_i\bar\psi_i\right)\psi_8\bar\psi_8\;.
\end{equation}
I will take $\As > 0$ and $\Bs < 0$. It is easiest to focus on the region $0 < |\Bs| \ll \As$. 
\\\\
First set $\Bs = 0$. At low energy an $SO(7)$-invariant fermion condensate will form:
\begin{equation}\label{eq:SO(7) condensate}
\bra i \psi_i\bar\psi_j \ket = \pm v\,\del_{ij}\;\;;\;\; i,j = 1,...,7\;\;\text{only}\;.
\end{equation}
This will induce an effective mass parameter for the first seven fermions: 
\begin{equation}\label{eq:m1}
m_1 = m_2 = ... = m_7 = \pm\, 14\As v\;.
\end{equation}
Upon turning on a small negative $\Bs$, one also induces a small mass parameter for the eighth fermion: 
\begin{equation}\label{eq:m8}
m_8 = \mp \,7|\Bs| v.
\end{equation}
In this region, the lowest-lying excitation above the ground state is this eighth fermion, and the mass gap of the theory is $|m_8|$. Let me emphasize that the parameters $m_1,...,m_8$ should not be confused with the $\eta\bar\eta$ mass parameter ``$m$" which is forbidden by the chiral $\zb_2$ symmetry in Eq.~(\ref{eq:physical Z2}).
\subsection{Absence of spontaneous symmetry breaking}\label{sec:no symmetry breaking}
In the introduction, I defined the bulk of an SPT phase to be invariant under a symmetry $G$ and to have a unique ground state. The goal of this section is to argue, purely within the low-energy field theory, that the $\zb_2$ symmetry of Eq.~(\ref{eq:physical Z2}) is not spontaneously broken. The goal of the next section will be to show that the ground state is non-degenerate.
\\\\
I showed in Sec.~\ref{sec:Z2} that the bilinears $i\psi_i\bar\psi_j$ are even under the physical chiral $\zb_2$ symmetry. In contrast, the bilinears in the original fermion fields, $i\eta_a\bar\eta_b$, are odd under the chiral $\zb_2$ symmetry (by definition). So the goal is first to argue that $\bra\eta_a\bar\eta_b\ket = 0$.
\\\\
Recall the relationship in Eq.~(\ref{eq:order disorder}) between $\eta\bar\eta$ and the $\psi$ order/disorder parameters, repeated below for convenience: 
\begin{equation}\label{eq:order disorder 2}
\sum_{a\,=\,1}^8 i\eta_a\bar\eta_a \propto \prod_{i\,=\,1}^8 \s_i^{(\psi)}+\prod_{i\,=\,1}^8\mu_i^{(\psi)} \;.
\end{equation}
In the $SO(8)$-invariant GN model, all values of the index $i = 1,...,8$ had to be interchangeable: for a fixed sign of the condensate, all Ising models were either ordered or disordered, and the expectation value $\sum_{a\,=\,1}^8\bra i\eta_a\bar\eta_a\ket$ was nonzero. By $SO(8)$ invariance, this means $\bra i\eta_a\bar\eta_a\ket \neq 0$ (no sum on $a$) for each $a = 1,...,8$.
\\\\
Now that the $SO(8)$ symmetry has been broken explicitly by a small negative $\Bs$, one has the situation described by Eqs.~(\ref{eq:m1}) and~(\ref{eq:m8}). When $\bra\sum_{i\,=\,1}^7i\psi_i\bar\psi_i\ket > 0$, one has $m_1 = ... = m_7 > 0$ and $m_8 < 0$. The first seven Ising models are ordered, but the eighth Ising model is \textit{disordered}:
\begin{align}
&\bra \s_1^{(\psi)}\ket = ... = \bra \s_7^{(\psi)}\ket \neq 0\;,\;\; \bra \s_8^{(\psi)}\ket = 0\;, \nonumber\\
&\bra \mu_1^{(\psi)}\ket = ... = \bra \mu_7^{(\psi)}\ket = 0\;,\;\; \bra \mu_8^{(\psi)}\ket \neq 0\;.
\end{align}
On the other hand, if $\bra\sum_{i\,=\,1}^7i\psi_i\bar\psi_i\ket < 0$, then $m_1 = ... = m_7 < 0$ and $m_8 > 0$. The first seven Ising models are disordered, but the eighth one is ordered:
\begin{align}
&\bra \s_1^{(\psi)}\ket = ... = \bra \s_7^{(\psi)}\ket = 0\;,\;\; \bra \s_8^{(\psi)}\ket \neq 0\;, \nonumber\\
&\bra \mu_1^{(\psi)}\ket = ... = \bra \mu_7^{(\psi)}\ket \neq 0\;,\;\; \bra \mu_8^{(\psi)}\ket = 0\;.
\end{align}
Either way, a small negative $\Bs$ allows the phase of the eighth Ising model to be \textit{anti}-correlated with the phase of the first seven, and one always concludes:
\begin{equation}
\left\bra \sum_{a\,=\,1}^8\eta_a\bar\eta_a\right\ket = 0\;.
\end{equation}
In the $SO(7)$ theory, the eight $\eta_a$ still transform as an 8-dimensional representation, so this implies $\bra \eta_a\bar\eta_a\ket = 0$ (no sum on $a$) for each $a = 1,..., 8$. 
\\\\
It is also necessary to check that $\bra \sum_{x\,=\,1}^8\chi_x\bar\chi_x\ket = 0$, since the $\chi$ mass bilinear is also odd under the $\zb_2$ operation. From the bosonization transformations, one finds the relation:
\begin{equation}\label{eq:order disorder 3}
\sum_{x\,=\,1}^8 i\chi_x\bar\chi_x \propto \left(\prod_{i\,=\,1}^8 \s_i^{(\psi)}-\prod_{i\,=\,1}^8\mu_i^{(\psi)} \right)\;.
\end{equation}
So the same argument will show that $\bra \sum_{x\,=\,1}^8\chi_x\bar\chi_x\ket = 0$ as well. Therefore, the physical $\zb_2$ symmetry remains unbroken at low energy.
\\\\
To emphasize the special nature of this particular model, now treat the original Majorana fermion fields, $\eta_1,...,\eta_8$, as Ising fermions, and consider adding and subtracting the analogs of Eqs.~(\ref{eq:order disorder 2}) and~(\ref{eq:order disorder 3}):
\begin{align}
&\sum_{i\,=\,1}^8i\psi_i\bar\psi_i+\sum_{x\,=\,1}^8 i\chi_x\bar\chi_x = \prod_{a\,=\,1}^8 \s_a^{(\eta)}\;,\nonumber\\
&\sum_{i\,=\,1}^8i\psi_i\bar\psi_i-\sum_{x\,=\,1}^8 i\chi_x\bar\chi_x = \prod_{a\,=\,1}^8 \mu_a^{(\eta)}\;.
\end{align}
Here I have dropped the unimportant overall numerical factor common to both equations. In the usual critical Ising model, one has a gapless theory whose Lagrangian is invariant under the exchange $\s_a^{(\eta)} \leftrightarrow \mu_a^{(\eta)}$. Here, however, one has $\bra \psi\bar\psi\ket \neq 0$, $\bra \chi\bar\chi\ket = 0$, and $\s_a^{(\eta)} = \s_b^{(\eta)}$, $\mu_a^{(\eta)} = \mu_b^{(\eta)}$ for all $a,b = 1,...,8$. (Remember now these are the order/disorder operators for the $\eta_a$ variables, which still transform as an 8-dimensional representation in the $SO(7)$ model.) 
\\\\
The $SO(7)$ model on the $``m = 0"$ manifold has a form of Kramers-Wannier invariance while still being a \textit{gapped} theory. This is very different from a garden-variety Ising model.
\subsection{Uniqueness of the ground state}\label{sec:degeneracy}
The ground state of the $SO(8)$ GN model is two-fold degenerate, as discussed in Sec.~\ref{sec:2fold degeneracy}. In the $SO(7)$ KF model, however, since $\bra \eta_a\bar\eta_b\ket = 0$, then the physical $\zb_2$ symmetry is unbroken and the ground state is unique. To determine the ground state, I study the potential for the KF model:
\begin{equation}
V(\psi,\bar\psi) = -\As\left( \sum_{i\,=\,1}^7 i\psi_i\bar\psi_i\right)^2-\Bs\left( \sum_{i\,=\,1}^7\psi_i\bar\psi_i\right)i\psi_8\bar\psi_8\;.
\end{equation}
The bosonization rules in Sec.~\ref{sec:triality} allow the bilinear of a single Majorana fermion to be expressed in terms of a non-chiral boson and its dual:
\begin{equation}
i\psi_{2I-1}\bar\psi_{2I-1} = \half\left[ \cos(2\pi\Ta_I)-\cos(2\pi\tilde\Ta_I)\right]\;,\;\; i\psi_{2I}\bar\psi_{2I} = \half\left[ \cos(2\pi\Ta_I)+\cos(2\pi\tilde\Ta_I)\right].
\end{equation}
For $\Bs = 0$, the potential is simply proportional to:
\begin{equation}
\left( \sum_{i\,=\,1}^7i\psi_i\bar\psi_i\right)^2\!\!\!\! = 2\!\sum_{I<J}^3 \cos(2\pi\Ta_I)\cos(2\pi\Ta_J)+\!\left[ \sum_{I\,=\,1}^3\cos(2\pi\Ta_I)\right]\!\!\left[ \cos(2\pi\Ta_4)-\cos(2\pi\tilde\Ta_4)\right].
\end{equation}
The distinction with respect to the GN potential [Eq.~(\ref{eq:GN potential})] is the contribution of three new terms involving $\cos(2\pi\tilde\Ta_4)$ and a relative factor of $\half$ in the $\cos(2\pi\Ta_4)$ terms. 
As before, minimization with respect to $\Ta_{1,2,3}$ implies $\sum_{I\,=\,1}^3\cos(2\pi\Ta_I) \neq 0$. Extremizing\footnote{Strictly speaking, just as it is not possible to simultaneously determine position and momentum, it is not possible to simultaneously determine $\Ta_4$ and $\tilde\Ta_4$. Nevertheless, the result obtained from this procedure is consistent with all expectations for the KF model, so I expect the result to be correct at least in some Gaussian sense. Perhaps the correct conclusion to draw from this exercise is that the formalism of abelian bosonization simply cannot capture this effect in the full quantum theory. I thank A. Kapustin and L. Fidkowski for discussions on this point.} with respect to $\Ta_4$ and the dual field $\tilde\Ta_4$ gives:
\begin{equation}\label{eq:minimization 2}
\left[ \sum_{I\,=\,1}^3\cos(2\pi\Ta_I)\right]\sin(2\pi\Ta_4) = 0\;,\;\;\left[ \sum_{I\,=\,1}^3\cos(2\pi\Ta_I)\right]\sin(2\pi\tilde\Ta_4) = 0\;.
\end{equation}
These conditions will only be satisfied if
\begin{equation}
\sin(2\pi\Ta_4) = \sin(2\pi\tilde\Ta_4) = 0\;.
\end{equation}
There are four logical possibilities:
\begin{align}
&\left. \begin{matrix}(1)\qquad\;\;\;\;\;\;\;\; \Ta_4,\,\tilde\Ta_4\in\zb \implies \cos(2\pi\Ta_4) = \cos(2\pi\tilde\Ta_4) = +1 \\ (2)\;\;\Ta_4-\half,\,\tilde\Ta_4-\half \in \zb \implies \cos(2\pi\Ta_4) = \cos(2\pi\tilde\Ta_4) = -1 \end{matrix} \right\} \implies \begin{matrix} i\psi_7\bar\psi_7 = 0\,,\\ i\psi_8\bar\psi_8 \neq 0\,. \end{matrix}\nonumber\\
&\nonumber\\
&\left. \begin{matrix} (3)\;\;\Ta_4\in\zb ,\,\tilde\Ta_4-\half\in\zb \implies \cos(2\pi\Ta_4) = +1,\,\cos(2\pi\tilde\Ta_4) = -1 \\ (4)\;\;\Ta_4-\half\in\zb,\,\tilde\Ta_4\in\zb \implies \cos(2\pi\Ta_4) = -1,\,\cos(2\pi\tilde\Ta_4) = +1 \end{matrix} \right\} \implies \begin{matrix} i\psi_7\bar\psi_7\neq 0\,,\\ i\psi_8\bar\psi_8 = 0\,. \end{matrix}
\end{align}
If $\cos(2\pi\Ta_I) \neq 0$ for $I \in\{ 1,2,3\}$, then $i\psi_i\bar\psi_i \neq 0$ for $i \in\{ 1,...,6\}$ (no implied sum). By $SO(7)$ symmetry, this implies $i\psi_7\bar\psi_7 \neq 0$. So the only consistent possibilities are options (3) and (4).
\\\\
I already argued that in the KF theory a simultaneous change in sign of all these cosines is a gauge symmetry. Therefore, there is only one ground state. In a fixed gauge, say choosing option (3) above, this ground state can be expressed as:
\begin{equation}\label{eq:ground state}
\Ta_I = (0,0,0,0)\,,\;\tilde\Ta_I = (0,0,0,\half)\;.
\end{equation}
In conclusion, a continuous tuning from $m > 0$ (topological phase) to $m < 0$ (trivial phase) does not pass through a point which breaks $\zb_2$, and it does not pass through a point for which the ground state is degenerate. This is consistent with the claim that there is no bulk phase transition between the two situations. It is also consistent with the corresponding study of the (0+1)$d$ fermions at the spatial boundaries, which would amount to a repeat of the lattice analysis in \cite{KF}.
\subsection{Propagator}
The standard Lehmann-K\"all\'en spectral decomposition\footnote{The notation in Eq.~(\ref{eq:LK}) is more or less standard. For a review, please see Appendix~\ref{sec:lehmann kallen}.} for the $\Ns_a$ propagator in a translationally invariant system is \cite{kallen, lehmann}:
\begin{align}\label{eq:LK}
\Ds^{\al\beta}_{ab}(p) &\equiv \int \!d^2x \;e^{\,ip\cdot x}i\bra 0|T\!\left(\Ns^\al_a(x)\bar \Ns^\beta_b(0)\right)\!|0\ket \nonumber\\
&= \frac{(-\!\!\not\! p+mI)^{\al\beta}}{p^2+m^2-i\e}\;\del_{ab}+\int_{m_{\text{th}}^2}^\infty \!\!d\tau\;\frac{(-\!\!\not\! p\,\rho^{(1)}_{ab}(\tau)+\tau^{1/2}\rho^{(2)}_{ab}(\tau)I)^{\al\beta}}{p^2+\tau-i\e}\;.
\end{align}
Here $\al,\beta$ are $SO(1,1)$ Dirac spinor indices, $m^2$ stands for the squared mass of the excitations for which the $\Ns_a(x)$ are good interpolating fields\footnote{By this I mean that if the state with one such excitation is denoted by $|1(p)\ket$, then the field $\Ns(x)$ has a well-defined matrix element $\bra 0|\Ns(x)|1(p)\ket = \bra 0|\Ns(0)|1(p)\ket\,e^{\,ip\cdot x}$, and the state $|1(p)\ket$ is responsible for the single particle pole at $p^2 = -m^2$ in the Lehmann-K\"all\'en decomposition of the propagator.}, $m_{\text{th}}$ is the threshold energy at which the multiparticle continuum begins. In the $SO(8)$-invariant Gross-Neveu model, the quantity $|m|$ is the $8_v$-fermion mass, which is the same as the kink mass (as required by triality).
\\\\
In the $SO(7)$-invariant KF model, it is clear that the quantity $|m|$ above should correspond to the rest energy of the excitation which creates a kink in the value of $\e_i \equiv \psi_i\bar\psi_i$, $i = 1,...,7$. Denote this rest energy by $m_{\text{kink}} \equiv |m|$. The $\Ns_a$ have the correct quantum numbers to annihilate those kinks, so the leading term in the expansion for $i\Ds(p)$ \textit{should} have an isolated single-particle pole at $p^2 = -m_{\text{kink}}^2$. 
\\\\
The situation of interest is when the mass term for $\Ns_a$ is absent in the Lagrangian. As emphasized previously, the low-energy theory remains invariant under the chiral $\zb_2$ transformation $\bar\Ns_a\Ns_b \to -\bar\Ns_a\Ns_b$: this symmetry is not spontaneously broken in the IR. The factor of $m$ in the single-particle contribution to the propagator would break this chiral $\zb_2$ invariance and therefore cannot appear in Eq.~(\ref{eq:LK}). 
\\\\
How can this apparent contradiction be reconciled? From an arithmetic point of view, the simplest resolution would simply be to cross out the $m$ in the numerator while keeping the denominator equal to $p^2+m_{\text{kink}}^2-i\e$. This peculiar prescription actually seems to be the correct answer. This requires careful consideration of the steps leading to the spectral decomposition in Eq.~(\ref{eq:LK}).
\\\\
The decomposition follows from inserting a resolution of the identity between the two fields $\Ns_a(x)$ and $\bar \Ns_b(0)$ in the definition of the propagator. The vacuum gives zero contribution. The single-particle states give the first nonvanishing contribution, which is proportional to $1/(p^2+m^2)$ for the appropriate choice of $m^2$. It is this single-particle contribution which requires further scrutiny.
\\\\
As remarked back in the introduction, in a (1+1)-dimensional field theory without spatial boundaries one normally thinks of the $m > 0$ Lagrangian and the $m < 0$ Lagrangian as equivalent descriptions of the same physical theory. One writes the Dirac Lagrangian as $\la = \bar\Psi (i\!\!\not\!\pa-mI)\Psi$ with a fixed sign of the mass parameter (say $m > 0$) and derives the equation of motion $(i\!\!\not\!\pa-mI)\Psi = 0$. (Here $\Psi$ is a generic Dirac spinor with no relation to the $\psi$ variables discussed previously.) This equation of motion is to be thought of as projecting out half of the degrees of freedom of the spinor $\Psi(x)$.
\\\\
But for a free theory on a compact space, the two choices of sign for $m$ define different phases. So it makes perfect sense that, if one is interested in the resolution of the identity in the form ``$1 = \sum_{\text{states}}|\text{state}\ket \bra \text{state}|$" for a particular phase, then one fixes a particular sign of $m$ and includes only that corresponding single-particle state in the sum. 
\\\\
For the special case of $n = 8k$ flavors, for example in the case $k = 1$ studied above, the whole point is that the $m > 0$ theory and the $m < 0$ theory \textit{can} be adiabatically deformed into each other without going through any phase transition. So in this model, in the resolution of the identity, one should \textit{sum over both possible signs of the mass parameter}. This removes the chirality violating term from the propagator without removing the pole at $p^2 = -m^2$ with $|m| = m_{\text{kink}}$.
\\\\
Therefore, the propagator for the $\Ns_a(x)$ fields on the ``$m = 0$" manifold of the $SO(7)$-invariant KF model should take the form:
\begin{equation}\label{eq:propagator}
\Ds_{ab}^{\al\beta}(p) = \frac{(-\!\!\not\! p)^{\al\beta}}{p^2+m_{\text{kink}}^2-i\e}\, \del_{ab}+\int_{m_{\text{th}}^2}^\infty d\tau\;\frac{(-\!\!\not\! p)^{\al\beta} \rho_{ab}(\tau)}{p^2+\tau-i\e}\;.
\end{equation}
This peculiar expression shows that the excitations described by $\Ns_a(x)$ propagate with an ordinary relativistic massive dispersion relation but nevertheless do not ever flip chirality. This is unfamiliar, but there is nothing wrong with it. 
\\\\
The amplitude for a left-handed fermion to flip chirality and turn into a right-moving fermion is still proportional to the mass parameter $m$, whose magnitude is $m_{\text{kink}}$. But in this case there is a doubling of the number of degrees of freedom, one on-shell fermion for each sign of the mass parameter, and the amplitude for a chirality flip is proportional to $m_{\text{kink}}+(-m_{\text{kink}}) = 0$. I refer to this phenomenon as ``parity doubling" in analogy with an effect in hadronic physics \cite{parity doubling}. (This term was also used by the authors of Ref.~\cite{mass without condensates 2}.)
\\\\
Along the ``$m = 0$" manifold between the ``trivial" and ``topological" phases of the free-fermion theory, indeed it is the case that the fermion propagator vanishes linearly with $p^{\,\mu}$ as $p^{\,\mu} \to 0$ \cite{gurarie propagator, cenke propagator}. The remaining issue is to identify the origin of the extra ``parity-conjugate" states.
\subsection{Parity doubling}\label{sec:parity doubling}
Before identifying the additional states in the KF model, let me pause briefly to discuss to what extent the ``parity doubling" effect is analogous to the one in hadronic physics. The parity transformation in 1+1 dimensions acts on a Dirac spinor\footnote{In this general discussion I use the standard notation $\psi$ for a Dirac spinor. The reader should not confuse this with the $8_+$ particles described previously.} $\psi$ as:
\begin{equation}
P:\qquad \psi(t,x) \to i\g^1\psi(t,-x)\;
\end{equation}
Since $\overline{\g^1} \equiv \g^0(\g^1)^\da \g^0 = +\g^1$, and since $(\g^1)^2 = -I$, the parity transformation flips the sign of the mass term:
\begin{equation}\label{eq:parity}
P:\qquad \bar\psi\psi \to -\bar\psi\psi\;.
\end{equation}
The free Dirac Lagrangian in 1+1 dimensions is typically considered to be invariant under parity because this sign flip can be compensated by a $\g^5$ field redefinition:
\begin{equation}\label{eq:field redef}
\zb_2:\qquad \psi(t,x) \to \g^5\psi(t,x)\;.
\end{equation}
But this is precisely the chiral $\zb_2$ transformation that has played such a crucial role in the previous arguments. In this interacting theory, I argue that we do not have the license to simply perform the field redefinition in Eq.~(\ref{eq:field redef}), and we should really think of the parity transformation in Eq.~(\ref{eq:parity}) as exchanging two different types of particles. It is for this reason that the term ``parity doubling" is an appropriate name for the interaction effect that results in the KF propagator.
\\\\
The parity doubling effect in hadronic physics in some ways is very similar to the effect studied in this paper, but in other important ways it is very different. (In the following I will have to assume that the reader has some familiarity with QCD. If not, the reader may feel free to take my choice of terminology at face value and proceed to Sec.~\ref{sec:kinks in KF}.) 
\\\\
For a long time there has been some qualitative evidence that baryons with the same transformation properties under flavor $SU(2)_L\x SU(2)_R$ but opposite eigenvalues of parity happen to have identical pole masses\footnote{In order for this to happen, the couplings for interaction terms which are invariant under $SU(2)_L\x SU(2)_R$ and couple these baryons to pions must be parametrically small \cite{restoration}. It is not a priori clear at all what dynamical mechanism should be responsible for this. The point is that the mass degeneracy of baryons with the same flavor quantum numbers but opposite parity cannot be explained solely by an effective restoration of $SU(2)_L\x SU(2)_R$.}. (See Ref.~\cite{parity doubling} for a detailed review and a more quantitative analysis of the data.) Since there are two types of fermions with exactly the same quantum numbers which are exchanged under parity\footnote{Let $B_+$ be the baryon for which parity $P$ acts as $P: B_+ \to +B_+$, and let $B_-$ be the baryon for which $P: B_- \to - B_-$. Then $P: \sq(B_+ + B_-) \to \sq(B_+ - B_-)$, and $P: \sq(B_+-B_-) \to \sq (B_+ + B_-)$, so the particles $B_1 \equiv \sq(B_+ + B_-)$ and $B_2 \equiv \sq(B_+ - B_-)$ are exchanged under parity. Under the parity doubling hypothesis, the pole mass of $B_1$ equals the pole mass of $B_2$.}, the energy spectrum is said to exhibit ``parity doubling" under this hypothesis. In this way, the effect in hadronic physics is completely analogous to the effect studied here. (Furthermore, it may be useful to note that just as the fermions in the KF model should be thought of as the kinks of the $\psi$ particles, the baryons in low energy QCD should also be thought of as solitons \cite{witten baryons}.)
\\\\
However, the two effects differ crucially in that the individual mass terms for the parity doublers in low energy QCD are \textit{not} forbidden by symmetry. Each type of particle has a propagator of the standard massive Dirac form, with an explicit mass term in the numerator (and the usual factor of $p^2+m^2$ in the denominator). In contrast, the symmetry which forbids the mass term in the KF model is \textit{conserved} at low energy, so each parity doubler cannot have a propagator of the standard massive Dirac form. This is of course exactly why I interpret the KF model as having a parity doubled spectrum in the first place, in order to consistently produce an isolated single particle pole at $p^2 = -m_{\text{kink}}^2$ without a term proportional to $m_{\text{kink}}$ in the numerator.
\\\\
This may be summarized as follows. In the KF model and in the low energy limit of QCD under the parity doubling hypothesis, each particle has a corresponding parity-conjugate particle with the same internal quantum numbers and the same pole mass. Therefore, both theories have a ``parity doubled" single particle spectrum. However, because of the very different symmetry requirements in the two theories, the fermion propagator in the KF model exhibits a zero, while the baryon propagators do not. The surprising feature of the KF model is that the fermions have mass without mass terms in the Lagrangian, while the surprising feature of a parity doubled spectrum in QCD is simply that the magnitudes of the baryon masses may be numerically equal.
\subsection{Kinks in KF}\label{sec:kinks in KF}
Now I will attempt to identify the extra particles in the KF model. Recall that in the $SO(8)$-invariant situation, the lowest-lying physical excitations transform as one of three distinct $8$-dimensional representations, namely $8_v$, $8_+$, or $8_-$. The explicit breaking of the symmetry to $SO(7)$ was defined by the decomposition
\begin{equation}
8_+ \to 7\oplus 1\;.
\end{equation}
Under this decomposition, the two other 8-dimensional representations \textit{remain} 8-dimensional representations:
\begin{align}
&8_v \to 8\;,\;\;8_- \to 8\;.
\end{align}
I intentionally do not distinguish between the two instances of ``8" above: the group $SO(7)$ has only one spinor representation. While the $8_v$ and $8_-$ were distinct in $SO(8)$, these degrees of freedom transform as the same representation of $SO(7)$ and therefore can mix in the low-energy theory. 
\\\\
One might worry that the nontrivial $\zb_2'$ charge of the $\chi$ variables [recall Eq.~(\ref{eq:gauge Z2})] might preclude this possibility. Another way to say this is that, in terms of the $\psi$ description, the theory contains ``even" kinks (the $\eta_a \sim 8_v$) and ``odd" kinks (the $\chi_x \sim 8_-$). But this $\zb_2'$ is broken (better to say ``Higgsed") by a nonzero condensate $\bra \psi_i\bar\psi_j\ket$. So, in a fixed gauge, one should be able to think of the $\eta$ particles and $\chi$ particles propagating together. The $\chi$ particles contribute the additional degrees of freedom required to realize the form in Eq.~(\ref{eq:propagator}) for the $\eta$ propagator. 
\\\\
Although this explicit identification of the appropriate states came from the study of a particular $1d$ model, it seems that this phenomenon should generalize to more complicated systems in higher dimensions. Consider a fermionic SPT phase classified by an integer $n$ whose classification can be reduced by interactions to $n \sim n+k$ for some $k$. My general conjecture is that the Hilbert space of the theory must be enlarged to include states corresponding to the opposite sign of the mass parameter. 
\\\\
If this is correct, then along the $``m = 0"$ manifold, the ``$m > 0$" fermions and ``$m < 0$" fermions should propagate together with a dispersion relation $p^2 = -m_*^2$ for some $m_* \neq 0$. The Green's function will be proportional to $\half[(\not \!\!p+m_*)+(\not\!\!p-m_*)] = \not \!\!p\;$ below the multiparticle threshold. This appears to be the only possibility that is consistent with all of the known results about Green's functions for symmetry protected topological phases. Unfortunately, I do not yet know how to check this proposal more explicitly.
\subsection{A remark about $\psi_8$}
In the previous sections, the KF model was studied for $0 \ll -\Bs \ll \As$. In the limit $\Bs = 0$, the Lagrangian describes one massless Majorana fermion ($\psi_8$), seven degenerate massive Majorana fermions ($\psi_{1},...,\psi_{7}$), and massive kinks. The form of the Lagrangian in this limit seems to indicate that the field $\psi_8$ is totally decoupled from $\psi_1,...,\psi_7$, so one might ask whether one could just ``delete" $\psi_8$ altogether and study the $SO(7)$ Gross-Neveu model \cite{BPS kinks}. 
\\\\
Instead of performing any detailed calculations, let me argue based on general principles that this cannot be done. The argument rests on the observation that the $\zb_2'$ transformation which flips the sign of $\psi_i\bar\psi_j$ is a gauge symmetry. 
\\\\
In 1+1 spacetime dimensions, it is possible to construct a Lagrangian formulation of this $\zb_2'$ gauge theory by embedding $\zb_2'$ into $U(1)'$ and writing a ``BF" theory with 0-form potential $B$ and 2-form field strength $F = dA$, where $A = A_\mu\, dx^\mu$ is the $U(1)'$ gauge potential. (For more details, see the already mentioned Ref.~\cite{coupling QFT to TQFT} as well as Refs.~\cite{seiberg instantons, sharpe 1, banks seiberg Zn, sharpe 2}.)
\\\\
The $\zb_2'$ transformation of interest is \textit{chiral}, in that it only rephases the left-moving fermions $\psi_i$ while leaving the right-moving fermions $\bar\psi_i$ unchanged. So the only way for this gauge theory to be non-anomalous is for the $U(1)'$ charges to sum to zero. For example, the fermions can be paired up as $\psi_{2I-1}+i\psi_{2I}$ and assigned the $U(1)'$ charges $(-1)^I$, with $I = 1,...,4$. 
\\\\
In Dirac notation,
\begin{equation}
F_I \equiv \sq\left[  \ml \bar\psi_{2I-1}\\ i\psi_{2I} \mr + i\ml \bar\psi_{2I}\\ i\psi_{2I} \mr\right]\;,
\end{equation}
the $U(1)'$ current for this embedding would be:
\begin{equation}
J^\mu = \sum_{I\,=\,1}^4 (-1)^I\; \overline F_I \g^\mu \half(I-\g^5) F_I\;.
\end{equation}
The field $\psi_8$ cannot be deleted from the Lagrangian, because an odd number of real fermions cannot all be charged under $U(1)'$. 
\section{Relation to electronic systems}
From the perspective of experimental condensed matter physics, the Kitaev-Fidkowski interaction may seem somewhat foreign, in that it singles out ``half" a fermionic degree of freedom. Although I do not propose any explicit experimental realization of this interaction, I do feel it would be useful to relate the model to another physical system with the same symmetries. 
\\\\
The basic required ingredients are four complex fermionic degrees of freedom. Two of these are already provided by spin, so one is interested in a problem with two degenerate ``channels," or ``flavors," of spinful fermions. 
\\\\
One such system is the two-channel Kondo effect \cite{affleck ludwig overscreened kondo, affleck ludwig multichannel}, where itinerant conduction electrons scatter off a two-state impurity localized at the origin. It turns out that this system is not quite appropriate, because it possesses only the $SO(5)\x SO(3)$ subgroup of $SO(8)$ instead of the larger $SO(7)$ subgroup. However, the one-channel two-impurity Kondo effect \cite{emery kivelson, affleck ludwig two impurities} is another system with the correct number of degrees of freedom, and in this model the global symmetry is exactly the desired $SO(7)$ subgroup which leaves a component of the $8_+$ fixed \cite{affleck ludwig jones so(7), maldacena so(8)}. 
\subsection{Two-channel, one-impurity Kondo effect}\label{sec:imp}
It is conceptually simplest to begin with the two-channel, one-impurity Kondo problem. One has two channels, or flavors, of conduction electrons in three spatial dimensions, labeled by an index $i = 1,2$. Each electron also has spin, labeled by $\al =\; \uparrow,\downarrow$. The impurity is taken to have spin-$\half$ and is localized at the origin. 
\\\\
The scattering of the conduction electrons on the impurity is dominated by the $\ell = 0$ angular momentum mode (``$s$-wave"), and hence can be reduced to a problem purely in the radial direction. Upon integrating over the angular variables, one is left with an effective 1+1 dimensional action on the half-line. In the low-energy theory, the residual effect of the impurity is to provide a boundary condition for the electronic degrees of freedom. 
\\\\
The complex left-handed fermion fields which describe the two flavors of electrons will be denoted as follows:
\begin{equation}\label{eq:electrons}
\ml e_{\uparrow 1}\\ e_{\downarrow 1}\\ e_{\uparrow 2}\\ e_{\downarrow 2} \mr \equiv \tfrac{1}{\sqrt2}\ml \eta_1+i\eta_2\\ \eta_3+i\eta_4\\ \eta_5+i\eta_6\\ \eta_7+i\eta_8 \mr\;.
\end{equation} 
There are $8\ox_A8 = 28$ different left-handed currents which generate infinitesimal $SO(8)_L$ transformations:
\begin{equation}
j_{ij} = i \psi_i \psi_j\;.
\end{equation}
In principle these $SO(8)_L$ currents can be written in terms of $\eta_a \sim 8_v$, $\psi_i \sim 8_+$, or $\chi_x \sim 8_-$. The corresponding expressions $j_{ab}$, $j_{ij}$, and $j_{xy}$ are related by a triality transformation. I have chosen to work in the $\psi_i \sim 8_+$ basis because, as will be seen shortly, if the physical conduction electrons are described as in Eq.~(\ref{eq:electrons}), then working in the $\psi_i \sim 8_+$ basis will effect a generalized ``spin-charge separation."
\\\\
Four of the $SO(8)_L$ currents, namely $j_{2I-1,I}$ $(I = 1,...,4)$, generate the four mutually commuting $U(1)_L$ subgroups of $SO(8)_L$. These define the four left-handed Cartan charges:
\begin{equation}
N_I = \int_{-\infty}^\infty dx\; i \psi_{2I-1}\psi_{2I}\;.
\end{equation}
Since $i\psi_{2I-1}\psi_{2I} = i\pa_x\ta_I$ and the $\ta_I$ are related to the $\ph_A$ in the same way as before, the $N_I$ can be expressed in terms of the conduction electrons as follows:
\begin{align}
N_I &= \half \sum_{A\,=\,1}^4\ml +&+&+&+\\ +&-&+&-\\ +&+&-&-\\ +&-&-&+ \mr_{IA}\int_{-\infty}^\infty dx\, i\eta_{2A-1}\eta_{2A} \nonumber\\
& = \int_{-\infty}^\infty dx\;\;\half\!\!\sum_{\al,\beta = \uparrow,\downarrow}\,\sum_{i,j\,=\,1,2} e_{\al i}^\da \ml \del_{\al\beta} \del_{ij}\\ \s^z_{\al\beta}\del_{ij}\\ \del_{\al\beta}\s^z_{ij}\\ \s^z_{\al\beta}\s^z_{ij} \mr e_{\beta j}\;.
\end{align}
From the definition in Eq.~(\ref{eq:electrons}), it is clear that $Q \equiv 2N_1$ measures the total electric charge, $S_z \equiv N_2$ measures the total $z$-component of spin, $F \equiv N_3$ measures the total ``flavor number," and $B \equiv N_4$ measures a fourth quantum number associated with the internal degrees of freedom of the impurity. 
\\\\
Since this problem turns out to be very closely related to the scattering of 3+1 dimensional relativistic fermions from an $SU(5)$ magnetic monopole \cite{rubakov, callan} (see also \cite{monopole rotor})  -- and in fact the two-impurity single-channel case is identical \cite{maldacena so(8)} -- I will take the liberty of calling this fourth quantum number ``baryon number." Thus for this problem the rotation to the $8_+$ basis describes the separation of charge, spin, flavor, and baryon number.
\\\\
The effect of the impurity is to impose the following boundary condition on the scattering of a left-moving fermion into a right-moving fermion at the physical boundary $x = 0$: \footnote{This is the location of the impurity; here $x \in [0,\infty)$ labels the \textit{radial} direction in the original $3d$ problem.}
\begin{equation}
\ml \psi_1\\ \psi_2\\ \psi_5\\ \psi_6\\ \psi_7 \mr \to +\ml \bar\psi_1\\ \bar\psi_2\\ \bar\psi_5\\ \bar\psi_6\\ \bar\psi_7 \mr\;,\;\; \ml \psi_3\\ \psi_4\\ \psi_8 \mr \to -\ml \bar\psi_3\\ \bar\psi_4\\ \bar\psi_8 \mr\;. 
\end{equation}
From this it is clear that the symmetry of the problem is reduced from $SO(8)$ to $SO(5)\x SO(3)$. 
\\\\
To make a connection with the Kitaev-Fidkowski model, the desired symmetry is the larger subgroup $SO(7)$, and in particular that $SO(7)$ which is defined by $8_+ \to 7\oplus 1$. It turns out that this is exactly the symmetry group for the one-channel, two-impurity problem. 
\subsection{One-channel, two-impurity Kondo effect}\label{sec:imp2}
In this case, one has only a single channel of physical conduction electrons, again labeled by spin $\al =\; \uparrow,\downarrow$. There are now two spin-$\half$ impurities, distributed symmetrically about the origin, say at locations $\vec x = \pm \half \vec R$ for some fixed constant vector $\vec R$. Because of this spatial separation of the two impurities, the long-distance description is one of a single effective spin-$1$ impurity, which couples differently to the different parities of the conduction electrons. Linear combinations of parity-even and parity-odd eigenstates provide the ``flavor" label, $i = 1,2$. 
\\\\
Again the long distance theory reduces to an effective 1+1 dimensional theory of 8 Majorana fermions with a boundary condition at $x = 0$. In this case, the boundary condition is:
\begin{equation}
\ml \psi_1\\ \psi_2\\ \psi_3\\ \psi_4\\ \psi_5\\ \psi_6\\ \psi_8 \mr \to +\ml \bar\psi_1\\ \bar\psi_2\\ \bar\psi_3\\ \bar\psi_4\\ \bar\psi_5\\ \bar\psi_6\\ \bar\psi_8 \mr\;,\;\; \psi_7 \to -\bar\psi_7\;.
\end{equation}
Hence the continuous global symmetry of the low-energy theory is $SO(7)$. Since these boundary conditions are written, intentionally, in terms of the $(\psi_i,\bar\psi_i)$ variables, indeed the \textit{correct} choice of $SO(7)$ subgroup is singled out. This is exactly what happens for the KF interaction (up to a trivial relabeling of $\psi_7 \leftrightarrow \psi_8$): the $U(1)$ rotations in the $(7,8)$-plane are explicitly broken by the $g'$ term, which singles out the 8th component of the $8_+$.
\\\\
To relate these two systems literally would require a physical implementation of the interactions in Eq.~(\ref{eq:KF}). In terms of electronic degrees of freedom on the lattice, these can arise from a Hubbard-Heisenberg interaction (see, for example, \cite{mass without condensates 1}). Writing the interaction in this manner has the advantage of being expressed in terms of familiar physical variables, but it has the disadvantage of obscuring the $SO(7)$ symmetry.
\\\\
It is enlightening to observe that, in terms of the $\chi_x \sim 8_-$ fermions, one has:
\begin{align}
N_I &= \half \sum_{X\,=\,1}^4\ml +&+&+&+\\ +&-&+&-\\ +&+&-&-\\ -&+&+&- \mr_{IX}\int_{-\infty}^\infty dx\, i\chi_{2X-1}\chi_{2X} \nonumber\\
&=\int_{-\infty}^\infty dx\;\; \half\!\!\sum_{\al,\beta = \uparrow,\downarrow}\,\sum_{i,j\,=\,1,2} \tilde e_{\al i}^\da \ml \del_{\al\beta} \del_{ij}\\ \s^z_{\al\beta}\del_{ij}\\ \del_{\al\beta}\s^z_{ij}\\ \s^z_{\al\beta}\s^z_{ij} \mr \tilde e_{\beta j}\;,
\end{align}
where I have defined new fields
\begin{equation}
\ml \tilde e_{\uparrow 1}\\ \tilde e_{\downarrow 1}\\ \tilde e_{\uparrow 2}\\ \tilde e_{\downarrow 2} \mr \equiv \sq\ml \chi_1+i\chi_2\\ \chi_3+i\chi_4\\ \chi_5+i\chi_6\\ \chi_7+i\chi_8 \mr\;.
\end{equation}
Evidently these new fields have the same charge, spin, and flavor quantum numbers as the original electron fields, but their baryon number is flipped: the $\tilde e_{\al i}$ are the ``antibaryons" of the $e_{\al i}$. So if $U(1)_B$ is broken by the interactions, then the $e_{\al i}$ and the $\tilde e_{\al i}$ will carry exactly the same quantum numbers. As discussed previously, these are the states which combine together to form a propagator proportional to $\not\! p/(p^2+m_*^2)$ for some $m_*^2 \neq 0$.
\section{Discussion}\label{sec:end}
In this paper I have emphasized a subtlety in the triality invariance of the $SO(8)$ Gross-Neveu model (Sec.~\ref{sec:GN}) and studied the $SO(7)$ Kitaev-Fidkowski model along the $``m = 0"$ manifold (Sec.~\ref{sec:KF}). The purpose was to obtain a more thorough understanding of the latter model in the continuum limit and to extract lessons for interacting relativistic quantum field theories in higher dimensions. 
\\\\
I pointed out that the two choices of sign for the $8_+$ condensate are gauge equivalent [Eq.~(\ref{eq:gauge Z2})], and hence the formation of this condensate does not indicate anything about the ground state degeneracy (Sec.~\ref{sec:degeneracy}). I also noted the important distinction between the $8_+$ and the $8_-$, and in particular the fact that only the $8_+$ bilinear is invariant under the physical chiral $\zb_2$ symmetry which emerges when the original mass parameter is set to zero [Eq.~(\ref{eq:physical Z2})]. 
\\\\
The main observation was that the fermion propagator should exhibit a form of ``parity doubling" for which states of equal and opposite mass parameters conspire to give a numerator proportional to $\g^\mu p_\mu$ while maintaining the single-particle pole at $p^2 = -m_{\text{kink}}^2 \neq 0$. This was motivated by the known spectrum of the $SO(8)$ GN model and the conclusion that the physical $\zb_2$ symmetry is not broken spontaneously.
\\\\
Since I cannot imagine any other possibility consistent with the known results for SPT phases as well as with the principles of relativistic quantum field theory, I conjecture that the fermion propagator in Eq.~(\ref{eq:propagator}) should describe the general situation: when a topological superconductor with $\zb$ classification can be reduced by interactions to some $\zb_k$, then the Hilbert space of the theory along the ``$m = 0$" must be doubled. It seems necessary to include states for both projections in the Dirac operator, one for a mass term $+m$ and one for a mass term $-m$. \textit{That is how a relativistic fermion can obtain mass without breaking any symmetry which forbids the mass terms in the Lagrangian.}
\\\\
I will conclude by proposing a novel application of the reduced classification of SPT phases to the study of elementary particle physics. It was already recognized that this phenomenon could be considered as a way to evade certain fermion doubling theorems on the lattice \cite{16fold}. However, there exists at least one example in which unwanted fermion doubling occurs purely within a field theoretic framework without any reference to a discretization of space. 
\\\\
In an attempt to unify not only the nuclear and electromagnetic forces in a grand unified $SO(10)$ theory \cite{fritzsch so(10), georgi so(10), lykken so(10)}, but also to combine the three generations of fundamental fermions into a single representation of a larger gauge group, a \textit{grand unified theory of families} was proposed based on the gauge group $SO(18)$ \cite{wilczek zee, bagger dimopoulos}. In this model, all known fermions could fit into a single 256-dimensional chiral spinor representation of $SO(18)$, and there was some hope that the peculiar repetitive family structure of the SM could be explained by group theory. 
\\\\
However, the desired property that spinors of $SO(2n+2m)$ contain spinors of $SO(2n)$ also proved to be the main phenomenological flaw of this approach: under the breaking of $SO(18)$ to $SO(10)$, the chiral spinor of $SO(18)$ splits into the desired families as well as ``mirror" families with the opposite quantum numbers (see, for example, \cite{zee mirror}). The problem was to explain why this mirror matter is not observed at low energy. 
\\\\
Since the study of interacting symmetry protected topological superconductors has suggested that the $SO(10)$ mirror fermions can likely be gapped out without generating a mass for the ordinary fermions, it is possible that this new insight from condensed matter theory may revive the $SO(18)$ model. This would be an interesting problem to work out in detail, but it is clearly beyond the scope of this paper.
\begin{center}\textit{Acknowledgements}\end{center}
This work was supported by NSF grants PHY07-57035 and PHY13-16748. I thank Alexei Kitaev, Joe Polchinski, Cenke Xu, Yonah Lemonik, Matthew Fisher, Andreas Ludwig, R. Shankar, Mark Srednicki, and A. Zee for very helpful discussions. 
\appendix
\section{Lattice Regularization}\label{sec:kitaev chain}
For the convenience of the quantum field theorist who is not necessarily familiar with the ultraviolet regularization described in the introduction, I will review briefly the $1d$ Kitaev chain \cite{kitaev chain}.
\subsection{Lattice}
Consider a $1d$ chain of sites indexed by $j,k = 1,...,2N$ with one real Majorana operator, $c_j$, per site:
\begin{equation}
\{c_j,c_k\} = \half \del_{jk}\;,\;\; c_j^\da = c_j\;.
\end{equation}
Introduce the following quadratic couplings between the fermions:
\begin{equation}\label{eq:J1J2}
H = -i\sum_{j\,=\,1}^{2N-1} \left\{ \half[1+(-1)^j]J_1+\half[1-(-1)^j]J_2\right\} c_j c_{j+1}\;.
\end{equation}
Define the sum and difference of $J_{1,2}$:
\begin{equation}
t \equiv J_1+J_2\;,\;\; m \equiv J_1-J_2\;.
\end{equation}
Then the Hamiltonian is simply:
\begin{equation}\label{eq:UV hamiltonian}
H = -\half i\sum_{j\,=\,1}^{2N-1} \left[ t+(-1)^j m\right]c_j c_{j+1}\;.
\end{equation}
In preparation for a change of basis into momentum space (Fourier transformation), it is best to rewrite each term in a symmetric fashion\footnote{The boundary terms will be taken care of later. The first part of the derivation will be for the continuum limit of the bulk of the chain.}:
\begin{equation}
\sum_j c_j c_{j+1} = \half \sum_j (c_j c_{j+1}-c_j c_{j-1})\;,\;\;\sum_j (-1)^j c_j c_{j+1} = \half \sum_j (-1)^j (c_j c_{j+1}+c_{j} c_{j-1})\;.
\end{equation}
\subsection{Fourier transform}
Let $a$ be the lattice spacing. With periodic boundary conditions, the wave at site $j$ is the same as the wave at site $2N+j$:
\begin{equation}
e^{\,ipaj} = e^{\,ipa(2N+j)} \implies e^{\,ipL} = 1\;,\;\; L \equiv 2Na\;.
\end{equation}
Therefore, the momentum is discrete and runs from $0$ to $2\pi/a$ [the ``fundamental region" or ``Brillouin zone" (BZ)]:
\begin{equation}
p = \frac{2\pi}{L}n\;,\;\; n = 0,1,...,\frac{L}{a}\;.
\end{equation}
The momentum space operators will be defined by:
\begin{equation}
c_j \equiv \sum_{p\,\in\,\text{BZ}}\,e^{\,ip aj}\,\tilde c_p\;.
\end{equation}
Plugging this into the Hamiltonian in Eq.~(\ref{eq:UV hamiltonian}) and symmetrizing appropriately gives:
\begin{equation}\label{eq:momentum space hamiltonian}
H = \half\sum_{p\,\in\,\text{BZ}}(\tilde c_{-p},\tilde c_{-(p+\pi/a)})\;h(p)\; \ml\tilde c_p\\ \tilde c_{p+\pi/a}\mr
\end{equation}
with the single particle Hamiltonian matrix
\begin{equation}\label{eq:h}
h(p) = t\sin(pa)\ml 1&0\\ 0&-1 \mr +m\cos(pa)\ml 0&-i\\ i&0 \mr \;.
\end{equation}
Squaring this gives the single particle dispersion relation:
\begin{equation}
h(p)^2 = E(p)^2I\;,\;\;E(p)^2 = t^2\sin^2(pa)+m^2\cos^2(pa)\;.
\end{equation}
Fill up the band up to the points at which the hopping term vanishes. In other words, define the Fermi momentum $p_F$ as the solutions to
\begin{equation}
\sin(p_F a) = 0\;.
\end{equation}
Since $p\in[0,2\pi/a]$, there are two distinguished points about which to linearize:
\begin{equation}
p_F = 0\;\;\text{ or }\;\;\frac{\pi}{a}\;.
\end{equation}
The goal is to describe fluctuations in the vicinity of both of these points. To do this, define
\begin{equation}
k \equiv p-p_F
\end{equation}
and expand the matrix in Eq.~(\ref{eq:h}) to linear order in $k$. The result is:
\begin{equation}
h(p = p_F+k) = \cos(p_Fa)\left(t\ml 1&0\\ 0&-1 \mr ka + m\ml 0&-i\\ i&0 \mr\right)+\op(ka)^2\;.
\end{equation}
Define the continuum fields by:
\begin{equation}
\tilde c_{\,p \,=\, p_F+k} \equiv a^{-1/2}\int dx\,e^{\,ikx}\,\eta_1(x)\;,\;\; \tilde c_{\,p+\pi/2\,=\,(p_F+\pi/2)+k} \equiv a^{-1/2}\int dx\,e^{\,ikx}\,\eta_2(x)\;,
\end{equation}
The form of the Hamiltonian in Eq.~(\ref{eq:momentum space hamiltonian}) already explicitly describes both points at which the hopping term intersects zero, so without loss of generality take $p_F = 0$ and hence $\cos(p_Fa) = +1$. The Hamiltonian in Eq.~(\ref{eq:momentum space hamiltonian}) becomes:
\begin{equation}
H \approx \int \! dx \sum_{a,b\,=\,1}^2\eta_a\left( t\, \s^z_{ab}\,i\pa_x +m\,a^{-1}\,\s^y_{ab}\right)\eta_b\;.
\end{equation}
After rescaling the fields into their canonical form and defining an appropriately rescaled mass parameter, one finds the standard Hamiltonian for a relativistic Majorana fermion
\begin{equation}
\Ns \equiv \ml \eta_2\\ i\eta_1 \mr\; \begin{matrix}  \leftarrow\text{ (right-moving)}\\ \leftarrow \text{ (left-moving) }\end{matrix}
\end{equation}
with mass
\begin{equation}
m = J_1-J_2\;.
\end{equation}
The corresponding Lagrangian is:
\begin{equation}
\la = \half \bar\Ns \left(i\!\!\not\!\pa-m\right)\Ns\;,\;\; \g^\mu = (\s_1,-i\s_2)\;.
\end{equation}
From this derivation it is clear that the ``$m > 0$" and ``$m < 0$" phases can be realized from the appropriate tuning of $J_2$ relative to $J_1$. In particular, the ``$m = 0$" manifold is realized when $J_2 = J_1$. 
\\\\
To determine which is the trivial phase and which is the topological phase, go back to the original lattice Hamiltonian in Eq.~(\ref{eq:J1J2}). If $J_2 \to 0$, then $c_1$ and $c_N$ become decoupled from the rest of the chain. The phase with $J_1 > J_2$ has an unpaired edge mode, while the phase with $J_1 < J_2$ does not. Therefore:
\begin{equation}
m < 0\;\;\text{is trivial}\;,\;\;m > 0\;\text{is topological}\;.
\end{equation}
\subsection{Time reversal}
In the lattice model, the peculiar time reversal transformation which squares to $+1$ (and is still antiunitary) is defined as
\begin{equation}
\zb_2^T:\qquad c_j \to (-1)^j c_j\;,\;\; i \to -i\;.
\end{equation}
The goal is to see how this transforms the continuum fields $\eta_a(t,x)$. The previous subsection showed that there are two distinguished points in momentum space, namely $p = 0$ and $p = \pi/a$. In the Fourier decomposition of the position space Majorana operators, this can be made explicit by writing:
\begin{equation}
c_j = \half\sum_{p\,\in\, \text{BZ}} e^{\,ipa j}\left[ \tilde c_{0+p}+(-1)^j\tilde c_{\frac{\pi}{a}+p}\right]\;.
\end{equation}
The Majorana operators at even and odd sites are:
\begin{align}
& c_{2J-1} = \sum_p e^{\,ipa(2J-1)}\left[ \tilde c_p+(-1)^{2J-1}\tilde c_{p+\pi/a}\right] = \sum_p e^{\,ipa(2J-1)}\left( \tilde c_p-\tilde c_{p+\pi/a}\right)\;,\\
&c_{2J} = \sum_p e^{\,ipa(2J)}\left[ \tilde c_p+(-1)^{2J}\tilde c_{p+\pi/a}\right] = \sum_p e^{\,ipa(2J)}\left(\tilde c_p+\tilde c_{p+\pi/a}\right)\;,
\end{align}
where $J = 1,...,N$. Since $\zb_2^T$ flips the sign of $c_{2J-1}$ but does not flip the sign of $c_{2J}$, it is clear that $\zb_2^T$ exchanges $\tilde c_p$ and $\tilde c_{p+\pi/a}$. 
\\\\
Therefore, in the continuum limit, time reversal acts as
\begin{equation}
\zb_2^T:\qquad \eta_a \to \s^x_{ab}\, \eta_b\;,\;\; i \to -i\;.
\end{equation}
In relativistic notation with a choice of gamma matrices $\g^\mu = (\s^x,-i\s^y)$ and $\g^5 = \g^0\g^1 = \s^z$, this becomes (up to an overall phase):
\begin{equation}
\zb_2^T:\qquad \Ns \to \g^0 \Ns\;,\;\; i \to -i\;.
\end{equation}
\section{Lehmann-K\"all\'en form of the propagator}\label{sec:lehmann kallen}
In this appendix I will review the steps that allow the fermion propagator to be expressed in the form of Eq.~(\ref{eq:propagator}). I will also briefly review the constraints of positivity on the spectral functions $\rho_{1,2}$ in order to assuage the reader that the unfamiliar form of the propagator in the KF model does not violate any theorems. In addition to the original papers \cite{kallen, lehmann}, the reader may also wish to consult the textbook by Itzykson and Zuber \cite{IZ}.
\subsection{Setup}
The Feynman propagator for a Dirac field $\psi(x)$ in the Poincar\'e representation $-p^2 = m^2$ in $D = d+1$ spacetime dimensions is defined as:
\begin{equation}
\Ds_{\al\beta}(x) \equiv i\ta(x^0)\bra 0|\psi_\al(x)\bar\psi_\beta(0)|0\ket -i\ta(-x^0)\bra 0|\bar\psi_\beta(0)\psi_\al(x)|0\ket \;.
\end{equation}
The state space is:
\begin{itemize}
\item vacuum: $|0\ket$
\item single particle state: $|1(p,s)\ket$, $p^2 = -m^2$, spin $s = \pm \half$
\item single antiparticle state: $|\bar 1(p,s)\ket$, $p^2 = -m^2$, spin $s = \pm \half$
\item multiparticle state: $|p,S,\xi\ket$, with some fixed value of $\tau \equiv -p^2 \geq m_{\text{th}}^2\geq m^2$ and some spin eigenvalue $S$. The ``threshold" scale $m_{\text{th}}$ defines the onset of the multiparticle continuum. All additional labels besides total momentum, spin, and $p^2$ are denoted collectively by $\xi$.
\end{itemize}
This defines the resolution of the identity operator:
\begin{align}\label{eq:1}
1 &= |0\ket\bra 0|+\int\frac{d^dp}{(2\pi)^d 2(\vec p^{\,2}+m^2)^{1/2}}\sum_{s\,=\,\pm}\left( |1(p,s)\ket\bra 1(p,s)| + |\bar 1(p,s)\ket \bra \bar 1(p,s)|\right) \nonumber\\
&+\int_{m_{\text{th}}^2}^\infty d\tau\int\frac{d^d p}{(2\pi)^d 2(\vec p^{\,2}+\tau)^{1/2}}\sum_S\,\intsum_\xi |p,S,\xi\ket \bra p,S,\xi|\,\del(p^2+\tau)
\end{align}
The delta function formally expresses the fact that the multiparticle state $|p,S,\xi\ket$ is in the Poincar\'e representation $-p^2 = \tau$.
\subsection{Wavefunctions}
Define the following single particle wavefunctions:
\begin{align}
&\bra 0|\psi_\al(x)|1(p,s)\ket \equiv u_\al(p,s)\,e^{\,ip\cdot x}\;,\;\; \bra \bar 1(p,s)|\psi_\al(x)|0\ket \equiv v_\al(p,s)\,e^{-ip\cdot x}\;.
\end{align}
These satisfy:
\begin{equation}\label{eq:spin sum}
\sum_{s\,=\,\pm \half}u_\al(p,s)\,\bar u_\beta(p,s) = -\!\!\not \!p_{\al\beta}+m\,\del_{\al\beta}\;,\;\; \sum_{s\,=\,\pm\half}v_\al(p,s)\,\bar v_\beta(p,s) = -\!\!\not \!p_{\al\beta}-m\,\del_{\al\beta}\;.
\end{equation}
For the multiparticle states, define the following wavefunctions:
\begin{align}
\bra 0|\psi_\al(x)|p,S,\xi\ket \equiv \As_\al(p,S,\xi)\,e^{\,ip\cdot x}\;,\;\; \bra p,S,\xi|\psi_\al(x)|0\ket \equiv \Bs_\al(p,S,\xi)\,e^{-ip\cdot x}\;.
\end{align}
By Lorentz invariance and parity, the scalar functions $\rho_1(\tau)$ and $\rho_2(\tau)$ can be defined by the following formula:
\begin{equation}\label{eq:M}
M_{\al\beta} \equiv \sum_S \intsum_\xi \As_\al(p,S,\xi)\bar\As_\beta(p,S,\xi)\,\del(p^2+\tau) \equiv -\!\!\not\! p_{\al\beta}\,\rho_1(\tau)+\tau^{1/2}\,\del_{\al\beta}\,\rho_2(\tau)\;,\;\;
\end{equation}
where $p^0 = (\vec p^{\,2}+\tau)^{1/2}$. Similarly, the scalar functions $\rho_1^c(\tau)$ and $\rho_2^c(\tau)$ (where the superscript $c$ is just part of the name of the function) can be defined by the formula:
\begin{equation}
\sum_S \intsum_\xi \Bs_\al(p,S,\xi)\bar\Bs_\beta(p,S,\xi) \equiv -\!\!\not\! p_{\al\beta}\,\rho_1^c(\tau)-\tau^{1/2}\,\del_{\al\beta}\,\rho_2^c(\tau)\;.
\end{equation}
The signs were chosen to match the analogous signs in Eq.~(\ref{eq:spin sum}). Invariance of the vacuum under charge conjugation implies:
\begin{equation}
\rho_1^c(\tau) = \rho_1(\tau)\;,\;\; \rho_2^c(\tau) = \rho_2(\tau)\;.
\end{equation}
\subsection{Result}
With Eq.~(\ref{eq:1}), the definitions in the previous subsection, and the relation
\begin{align}
\int\frac{d^D p}{(2\pi)^D}&\frac{-i}{p^2+m^2-i\e}\,e^{\,ip\cdot x}\,f(p) \nonumber\\
&= \int \frac{d^d p}{(2\pi)^d 2(\vec p^{\,2}+m^2)^{1/2}}\left(\ta(x^0)\,e^{\,ip\cdot x}\, f(p)+\ta(-x^0)\,e^{-ip\cdot x}\, f(-p)\right)\;,
\end{align}
the Lehmann-K\"all\'en form is obtained:
\begin{equation}\label{eq:lehmann kallen}
\Ds_{\al\beta}(x) = \int \frac{d^D p}{(2\pi)^D}\,e^{\,ip\cdot x}\left( \frac{-\!\!\not \! p_{\al\beta}+m\,\del_{\al\beta}}{p^2+m^2-i\e} + \int_{m_{\text{th}}^2}^\infty d\tau\;\frac{-\!\!\not\! p_{\al\beta}\,\rho_1(\tau)+\tau^{1/2}\,\del_{\al\beta}\,\rho_2(\tau)}{p^2+\tau-i\e}\right)\;.
\end{equation}
The additional flavor labels in Eq.~(\ref{eq:propagator}) present no additional complication and can simply be added on according to the invariance requirements of the appropriate flavor symmetry group.
\subsection{Positivity constraints}
The functions $\rho_1(\tau)$ and $\rho_2(\tau)$ satisfy certain inequalities as a result of positivity.  Recall the matrix $M$ defined in Eq.~(\ref{eq:M}). Multiplying on the right by $\g^0$ and taking the trace gives:
\begin{align}
\tr(M\g^0) &= \sum_S \intsum_\xi \sum_{\al}\left|\bra 0|\psi_\al(0)|p,S,\xi\ket\right|^2\,\del(p^2+\tau) = \tr(I) (\vec p^{\,2}+\tau)^{1/2}\,\rho_1(\tau)\;.
\end{align}
Since $\left|\bra 0|\psi_\al(0)|p,S,\xi\ket\right|^2 \geq 0$, the first positivity constraint is:
\begin{equation}
\rho_1(\tau) \geq 0\;.
\end{equation}
Similarly, but with a few more intermediate steps, multiplying the quantity $(\g^\mu p_\mu-\tau^{1/2}I)M(\g^\mu p_\mu-\tau^{1/2}I)$ on the right by $\g^0$ and taking the trace gives:
\begin{align}
&\tr\left( (\not\! p-\tau^{1/2}I)M(\not\! p-\tau^{1/2}I)\g^0\right) \nonumber\\
&\qquad = \sum_S \intsum_\xi \sum_{\al}\left| \bra 0|\left[ (i\!\!\not\!\pa+\tau^{1/2}I)\psi(x)\right]_\al|p,S,\xi\ket\right|^2\,\del(p^2+\tau) \nonumber\\
&\qquad = 2\tau(\vec p^{\,2}+\tau)^{1/2}\,\tr(I)\,(\rho_1(\tau)+\rho_2(\tau))\;.
\end{align}
Since $\left| \bra 0|\left[ (i\!\!\not\!\pa+\tau^{1/2}I)\psi(x)\right]_\al|p,S,\xi\ket\right|^2 \geq 0$, the second positivity constraint is:
\begin{equation}
\rho_1(\tau)+\rho_2(\tau) \geq 0\;.
\end{equation}
As defined, the scalar function $\rho_2(\tau)$ can have either sign. So really this second constraint amounts to:
\begin{equation}
\rho_1(\tau) \geq |\rho_2(\tau)|\;.
\end{equation}
Although it may be unfamiliar, it is internally consistent to have $\rho_2 = 0$ even with massive poles in the propagator.

\end{document}